%% file: verticalstructure.tex
\documentclass[useAMS,usenatbib]{mn2e}

\usepackage{times}

\usepackage{subfigure}

\usepackage{amsmath}
\usepackage{amssymb}
\usepackage{graphicx}
\usepackage{latexsym}
\usepackage{aas_macros}

\usepackage{url}
\usepackage{epstopdf}

\usepackage[squaren]{SIunits}

\input{commands}

\input{planetcommands}
\input{myunits}

\newcommand{\tset}[1]{\ensuremath{t_{\text{sett}}^{#1}}}
\newcommand{\tstir}[1]{\ensuremath{t_{\text{stir}}^{#1}}}

\newcommand{\Machz}[1]{\ensuremath{\mathcal{M}_{\text{z}}^{#1}}}

\begin{document}

\title[Vertical Gap Structure]{The Vertical Structure of Planet-induced Gaps in Proto-Planetary Discs}
\author[R. G. Edgar and A. C. Quillen]{Richard G. Edgar\thanks{rge21@pas.rochester.edu} and Alice C. Quillen\thanks{aquillen@pas.rochester.edu}  \\
Department of Physics and Astronomy, University of Rochester, Rochester, NY 14627} 
\date{\today}

\pagerange {\pageref{firstpage}--\pageref{lastpage}}

\label{firstpage}


\maketitle

\begin{abstract}
Giant planets embedded in circumstellar discs are expected to open gaps in these discs.
We examine the vertical structure of the gap edges.
We find that the planet excites spiral arms with significant (Mach number of a half) vertical motion of the gas, and discuss the implications of these motions.
In particular, the spiral arms will induce strong vertical stirring of the dust,  making the edge appeared `puffed up' relative to the bulk of the disc.
Infra-red observations (sensitive to dust) would be dominated by the light from the thick inner edge of the disc.
Sub-millimetre observations (sensitive to gas velocities) would appear to be hot in `turbulent' motions (actually the ordered motion caused by the passage of the spiral arms), but cold in chemistry.
Resolved sub-millimetre maps of circumstellar discs might even be able to detect the spiral arms directly.
\end{abstract}


\begin{keywords}
planetary systems : protoplanetary disks
\end{keywords}


\input{intro}

\input{numeric}

\input{results}

\input{discuss}

\input{conclude}


\appendix

\input{changegamma}

\input{viscouscomponents}

\input{resolution}


\bibliography{general}
\bibliographystyle{astron}


\section*{Acknowledgements}

The authors acknowledge support from NSF grants AST-0406799, AST-0098442, AST-0406823, and NASA grants ATP04-0000-0016 and NNG04GM12G (issued through the Origins of Solar Systems Program).
This work is supported in part by the U.S. Department of Energy under Grant No. B523820 to the Center for Astrophysical Thermonuclear Flashes at the University of Chicago.
The computational results presented here were performed on the facilities of the National Center for Supercomputing Applications and the Texas Advanced Computing Center, using TeraGrid time grant TG-AST070026.
RGE is also most grateful to Artur Gawryszczak for help with persuading \textsc{Flash} to work in polar co-ordinates.

\bsp

\label{lastpage}

\end{document}

%% file: commands.tex
\newcommand{\scinot}[2]{\ensuremath{#1 \times 10^{#2}}}

\newcommand{\paren}[1]{\left ( #1 \right )}

\newcommand{\parenfrac}[2]{\paren{\frac{#1}{#2}}}

\newcommand{\differd}[1]{\textrm{d}^{#1}}
\newcommand{\differ}[1]{\differd{}#1}
\newcommand{\sdiffer}[1]{\, \differ{#1}}



%% file: planetcommands.tex
\newcommand{\OmegaKep}[1]{\ensuremath{\Omega_{\text{kep}}^{#1}}}

\newcommand{\OmegaPlanet}[1]{\ensuremath{\Omega_{\text{p}}^{#1}}}

\newcommand{\rHill}[1]{\ensuremath{r_{\text{H}}^{#1}}}

%% file: myunits.tex
\addunit{\cm}{\centi\metre}

\addunit{\dyne}{dyn}
\addunit{\erg}{erg}

\addunit{\gauss}{gauss}

\addunit{\persqcm}{\per \cm \squared}
\addunit{\persqcmnp}{\cm \rpsquared}

\addunit{\percubiccm}{\per \cm \cubed}
\addunit{\percubiccmnp}{\cm \rpcubed}

\addunit{\grampersqcm}{\gram \persqcm}
\addunit{\grampersqcmnp}{\gram \usk \persqcmnp}
\addunit{\grampercubiccm}{\gram \percubiccm}
\addunit{\grampercubiccmnp}{\gram \usk \percubiccmnp}

\addunit{\ergpercubiccm}{\erg \percubiccm}
\addunit{\ergpercubiccmnp}{\erg \usk \percubiccmnp}

\addunit{\molpercubiccm}{\mole \percubiccm}
\addunit{\molpercubiccmnp}{\mole \usk \percubiccmnp}

\addunit{\cmpersec}{\cm \per \second}
\addunit{\cmpersecnp}{\cm \usk \reciprocal \second}
\addunit{\kmpersecnp}{\kilo\metre\usk\reciprocal\second}

\addunit{\cmpersecsq}{\cm \per \second \squared}
\addunit{\cmpersecsqnp}{\cm \usk \second \rpsquared}

\addunit{\gramcmpersec}{\gram \usk \cmpersec}
\addunit{\gramcmpersecnp}{\gram \usk \cmpersecnp}

\addunit{\gramsqcmpersec}{\gram \usk \cm \squared \per \second}
\addunit{\gramsqcmpersecnp}{\gram \usk \cm \squared \second \rpsquared}

\addunit{\yyear}{yr} 
\addunit{\MBU}{MBU} 

\addunit{\jansky}{Jy}

\addunit{\magnitude}{mag}

\addunit{\cmsqpergramnp}{\centi\metre\squared\usk\reciprocal\gram}

\newcommand{\mySun}{\odot}

\addunit{\Msol}{\ensuremath{\mathrm{M}_{\mySun}}}
\addunit{\Rsol}{\ensuremath{\mathrm{R}_{\mySun}}}
\addunit{\Lsol}{\ensuremath{\mathrm{L}_{\mySun}}}
\addunit{\Zsol}{\ensuremath{\mathrm{Z}_{\mySun}}}

\providecommand{\earth}{\oplus}

\addunit{\Mearth}{\ensuremath{\mathrm{M}_{\earth}}}
\addunit{\Rearth}{\ensuremath{\mathrm{R}_{\earth}}}

\addunit{\Mjup}{\ensuremath{\mathrm{M}_{J}}}
\addunit{\Rjup}{\ensuremath{\mathrm{R}_{J}}}

\addunit{\AU}{au}
\addunit{\lightyear}{ly}
\addunit{\parsec}{pc}

%% file: intro.tex
\section{Introduction}
\label{sec:intro}

Recently, \citet{2005ApJ...621..461D,2005ApJ...630L.185C} identified three young stars (CoKu Tau/4, GM Aur and DM Tau) where the circumstellar disc appears to have an inner hole.
These observations were performed with the \emph{Spitzer} Space Telescope.
CoKuTau/4 shows no evidence of continued accretion, while GM Aur and DM Tau are still accreting.
GM Aur even shows some evidence for an optically thin disc in the innermost portions of the hole.
These systems are often referred to as `transition discs.'
It has been suggested that these holes were cleared by a planet \citep{2004ApJ...612L.137Q,2004ApJ...612.1152V,2006ApJ...640.1110V}.

In this paper we shall discuss the \textsc{3d} structure of gaps created in discs by embedded planets, and the possible observational consequences.
This is important, since \emph{Spitzer} cannot image these systems directly; we are obliged to analyse spectral energy distributions (SEDs).
A key problem in interpreting \emph{Spitzer} measurements is our lack of knowledge about the dust and gas dynamics close to a planet-induced disc gap.
In particular, the disc edges appear to have $h/r \approx 0.1$, which is too large for simple (two dimensional) thin disc models to be wholly appropriate.

Despite these limitations, \citet{2006MNRAS.373.1619R} have already demonstrated (using \textsc{2d} calculations) the interesting possibility of a `dust filter' acting close to the edge.
This could be a way of permitting GM Aur and DM Tau to continue to accrete \citep[cf][ below]{2006ApJ...641..526L}, while still retaining their disc holes.
Several authors \citep{2004A&A...425L...9P,2007Ap&SS.311....3M,2007A&A...474.1037F} have noted that a gap might open in the dust, before the gas disc shows similar behaviour.
This is due to the dust disc having zero viscosity.
Vertical settling of the dust within the disc \citep{2006ApJ...638..314D} will also be relevant.
If any cross-gap flow is primarily from the upper layers of the disc, then it will naturally be depleted in dust.

Most previous numerical work on protoplanet forming discs has been performed in \textsc{2d} \citep[see][ for a selection of codes which have been used to study the problem]{2006MNRAS.370..529D}, due to the large computational cost of \textsc{3d} calculations.
Most previous work in \textsc{3d} has concentrated on migration \citep[e.g.][]{2001ApJ...547..457K,2002A&A...385..647D}, and the flow in the circumplanetary region \citep{2006A&A...445..747K}.
\citet{2003MNRAS.341..213B} studied migration rates of planets in \textsc{3d} discs, finding that \textsc{3d} considerations slowed Type I migration rates \citep[as predicted by][]{2002ApJ...565.1257T}.
\citet{2006A&A...459L..17P} studied the effect of including radiative transfer in their \textsc{3d} code, concluding that the pressure gradients induced could reverse Type I migration.
These studies have illustrated that \textsc{2d} calculations are poor approximations to the true flow in the vicinity of the planet.
Of course, for global calculations of the effect of lower mass planets on a disc, the \textsc{2d} approximation is more than adequate (and we shall see an aspect of this in our results below).
In this paper, we shall demonstrate important effects in the gap edges which are only seen in \textsc{3d} calculations.

\citet{2005ASPC..341..839B,2006ApJ...641..534B} studied the vertical structure of shocks in self-gravitating discs, without an embedded protoplanet
They particularly examined the behaviour of the gas before and after the shock jump.
Depending on the nature of the equation of state and the degree to which self gravity was important, they found that the disc might compress or expand vertically after passing a shock.
They find that a breaking wave could be excited on the surface layers of the disc, which has the potential to mix the disc radially.
In this paper, we discuss the effect of the spiral arms raised by an embedded planet.

In this paper, we discuss our numerical method in Section~\ref{sec:numerics}.
We present our results in Section~\ref{sec:results}, and discuss their consequences in Section~\ref{sec:discuss}.
Our conclusions are presented in Section~\ref{sec:conclude}.

%% file: numeric.tex
\section{Numerical Setup}
\label{sec:numerics}

In this section, we shall describe the code we used in our numerical experiments.
Full \textsc{3d} models of circumstellar discs are extremely challenging, requiring over an order of magnitude more computing power than \textsc{2d} models of comparable resolution.

Our code is based on the \textsc{Flash} code of \citet{2000ApJS..131..273F}, an adaptive mesh refinement (AMR) code based around a Piecewise-Parabolic Method (PPM) hydrodynamics solver.\footnote{The source code is available at \url{http://flash.uchicago.edu/}}
We made several modifications to the code, to adapt it for these experiments.
First, we adjusted the transport step to enforce conservation of angular momentum about the $z$-axis \citep{1998A&A...338L..37K}.
The detailed implentation of this is that used by the polar \textsc{Flash} code in the comparison of hydrodynamics codes presented by \citet{2006MNRAS.370..529D}.
We have verified that the $z$ component of the angular momentum is conserved to machine precision.
Our discs were isothermal in the azimuthal and vertical directions, and had a temperature gradient in the radial direction.
The pressure at any point in the disc is calculated from this temperature profile and the local density, providing the input states to the Riemann solver.
In solving the Riemann problem, the solver assumes that the gas has $\gamma=1.1$.
The radial temperature gradient is set to produce a flaring disc ($h/r \propto r^a$ for some $a$).
We added a simple \textsc{nbody} solver, to model the presence of a planetary system.
In this paper, we only use two bodies (a star and a planet), and they do not feel the gravitational effect of the disc.
Consequently, the orbit of the planet did not evolve in runs presented herein.
In addition, we added physical viscosity to the code.
We implemented all components of the viscous stress tensor, but the calculations shown here only use the $r$--$\phi$ component, since it is not clear that the true effects of MHD turbulence can be reproduced by a physical viscosity \citep[see][]{2003ApJ...589..543W}.
Buffer zones in the radial direction damp waves, preventing them reaching the inner or outer boundaries.
The buffer zones are implemented in the manner of \citet{2006MNRAS.370..529D}.

Our computational grid ranges between $r=\unit{\scinot{2}{13}}{\centi\metre}$ and \unit{\scinot{2}{14}}{\centi\metre}, between $z=0$ and \unit{\scinot{3}{13}}{\centi\metre} vertically and over a full circle in azimuth.
The buffer zones reduce the effective radial extent to $\unit{\scinot{3.2}{13}}{\cm} < r < \unit{\scinot{1.7}{14}}{\cm}$.
We used a uniform mesh, with $(n_r, n_{\phi}, n_z) = ( 128, 384, 64 )$.
This makes the grid cells almost square around the planet in the $r$--$\phi$ plane, but with higher resolution in the $z$ direction.
The increased $z$ resolution is necessary to avoid under-resolving the vertical structure of the inner disc.
We do not use the AMR capabilities of \textsc{Flash}, since we are not certain that the jumps in refinement do not produce any artifacts.
The edge of the gap will always lie roughly parallel to the fine-coarse interfaces, opening the possibility that small errors in the interpolation between levels will build up over time \citep[see also the comments of][]{2007A&A...470..367C}.
Our planetary system consisted of a planet in a fixed circular orbit around a \unit{1}{\Msol} star.
We performed runs with planetary masses of $1$ and \unit{2}{\Mjup}.
The orbital semi-major axis was $\unit{\scinot{7.5}{13}}{\centi\metre} = \unit{5}{\AU}$.

The initial density profile of the disc was constructed so that the surface density was constant at $\Sigma_{0} = \unit{\ensuremath{10^{3}}}{\grampersqcmnp}$.
There is great uncertainty in the appropriate density profile of protoplanetary discs, with estimates ranging between the constant value we use here, and the $\Sigma \propto r^{-3/2}$ form of the Minimum Mass Solar Nebula.
The disc was mildly flaring, with
\begin{equation}
\frac{h}{r} = 0.07 \parenfrac{r}{r_0}^{0.25}
\label{eq:flaringsetup}
\end{equation}
with $r_0 = \unit{\scinot{1.5}{14}}{\centi\metre}$.
We made use of equation~4 of \citet{2002ApJ...565.1257T} to adjust the initial orbital velocity of the gas to take account of the radial pressure gradient.

We neglect self-gravity in our calculations.
Together with our assumption of a vertically isothermal disc, this prevents us reproducing much of the behaviour observed by \citet{2005ASPC..341..839B,2006ApJ...641..534B}.
However, since our discs contain a planet, self-gravity is not necessary to generate spiral arms.
The assumption of a vertical isothermality also supresses the `wave channelling' behaviour described by \citet{2002MNRAS.332..575B}, where spiral waves refract towards the surface of a vertically stratified disc.
Nevertheless, these calculations provide a useful starting point for analysing the vertical structure of gap edges.

%% file: results.tex
\section{Results}
\label{sec:results}

We shall present the results of our numerical experiments, grouped by the planet mass.
For each planet mass, we performed runs with different values for the coefficient of viscosity, $\nu$.
The first run set $\nu=0$, and hence is dependent solely on the intrinsic numerical dissipation of the code.\footnote{Note that there is no reason to expect the numerical dissipation to act like a viscous term}
Two non-zero values were tested, $\nu = \unit{\ensuremath{10^{15}}}{\cm \squared \usk \reciprocal \second}$ and $\nu = \unit{\ensuremath{10^{16}}}{\cm \squared \usk \reciprocal \second}$.
We shall refer to these runs are `zero,' `low' and `high' viscosity respectively.

The high and low viscosity discs have Reynolds numbers at the planet's orbit of $\mathcal{R} \equiv r^2 \Omega / \nu = 10^5$ and~$10^4$ respectively.
There are a number of different criteria for determining whether a planet should open a gap.
First is the tidal criterion of \citet{1993prpl.conf..749L}:
\begin{equation}
q > 3 \paren{h/r}^3
\end{equation}
This is obtained by comparing the size of the Hill sphere, $\rHill{} \equiv r (q/3)^{1/3}$, which is the volume over which the planet's gravity dominates, to the thickness of the disc.
More commonly used is the viscous condition
\begin{equation}
q > 40 \mathcal{R}^{-1}
\end{equation}
as discussed by \citet{1999ApJ...514..344B}.
\citet{2006Icar..181..587C} combined these two conditions into a single one:
\begin{equation}
\frac{3}{4} \cdot \frac{h}{\rHill{}}+\frac{50}{q \mathcal{R}} \leq 1 
\end{equation}
In Table~\ref{tbl:gapopenpredict} we show the gap-opening predictions for each criteria, listed by planet mass and disc viscosity.
There is significant uncertainty, underlined by the fact that the definition of a gap is somewhat arbitrary \citep[see][]{2007MNRAS.378..966H}.

\begin{table}
\centering
\begin{tabular}{c|ccc}
Criterion  & \unit{1}{\Mjup} & \unit{2}{\Mjup} \\
\hline
Tidal      & ggg & ggg \\
Viscous    & ggn & ggn \\
Combined   & gnn & ggn
\end{tabular}
\caption{Predictions of gap opening according to different criteria. The criteria are explained in the text. Each entry lists three values, for the zero, low and high viscosity cases respectively. A `n' indicates that no gap should form, a `g' that a gap is expected}
\label{tbl:gapopenpredict}
\end{table}

We shall study the state of the disc after 100 orbits.
Although a steady state is never reached \citep[see, e.g.][]{2006ApJ...640.1110V,2007ApJ...663.1325E}, the condition of the disc after 100 orbits is generally accepted as being representative \citep{2006MNRAS.370..529D}.
Each orbit requires approximately \unit{300}{\hour} of CPU time on NCSA's \texttt{mercury} cluster.


\subsection{Jupiter Mass Planet}

\begin{figure}
\centering
\subfigure{\includegraphics[scale=0.6]{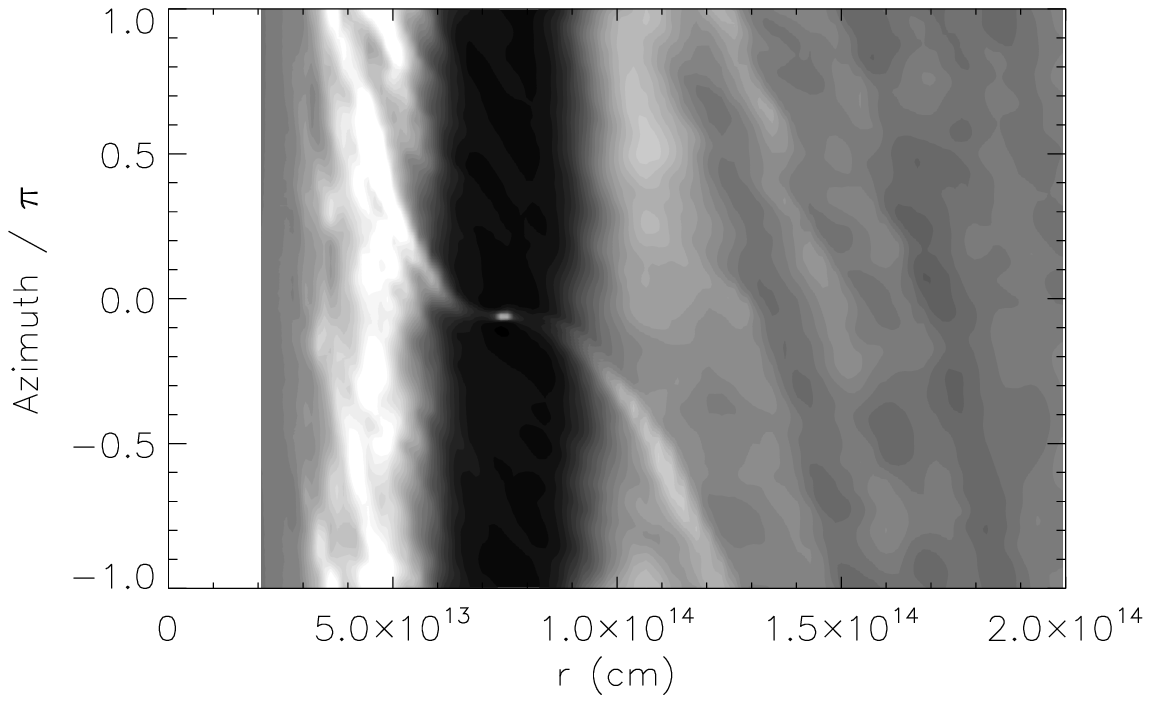}}
\subfigure{\includegraphics[scale=0.6]{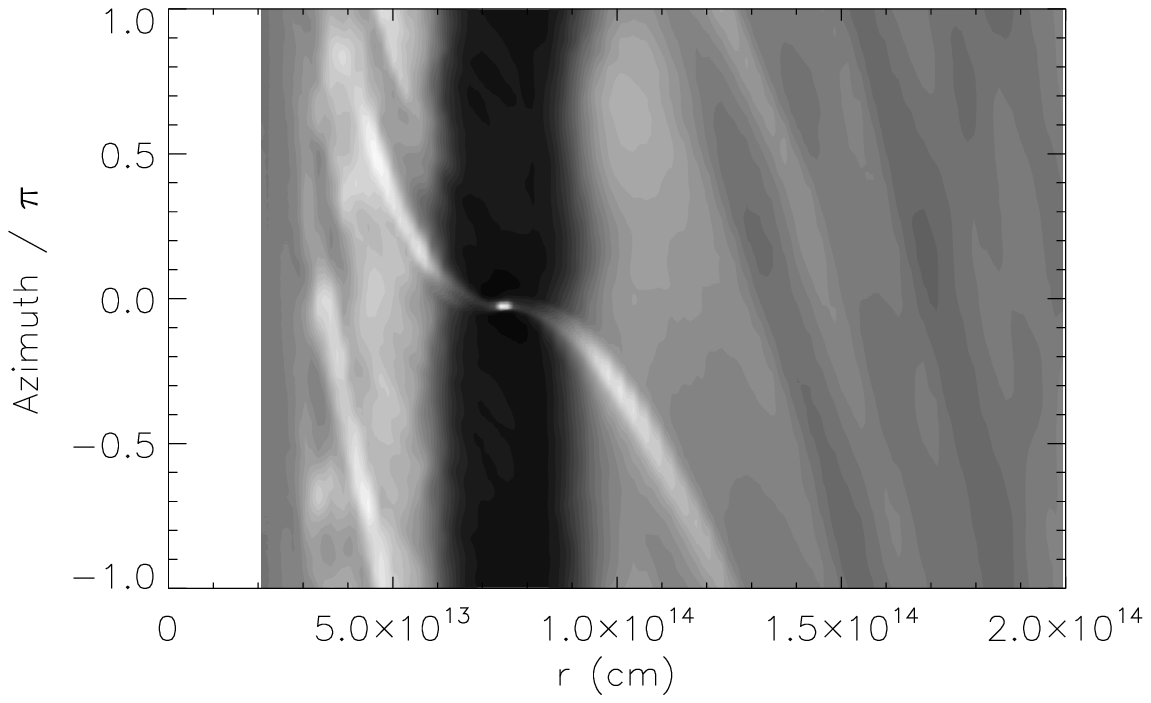}}
\subfigure{\includegraphics[scale=0.6]{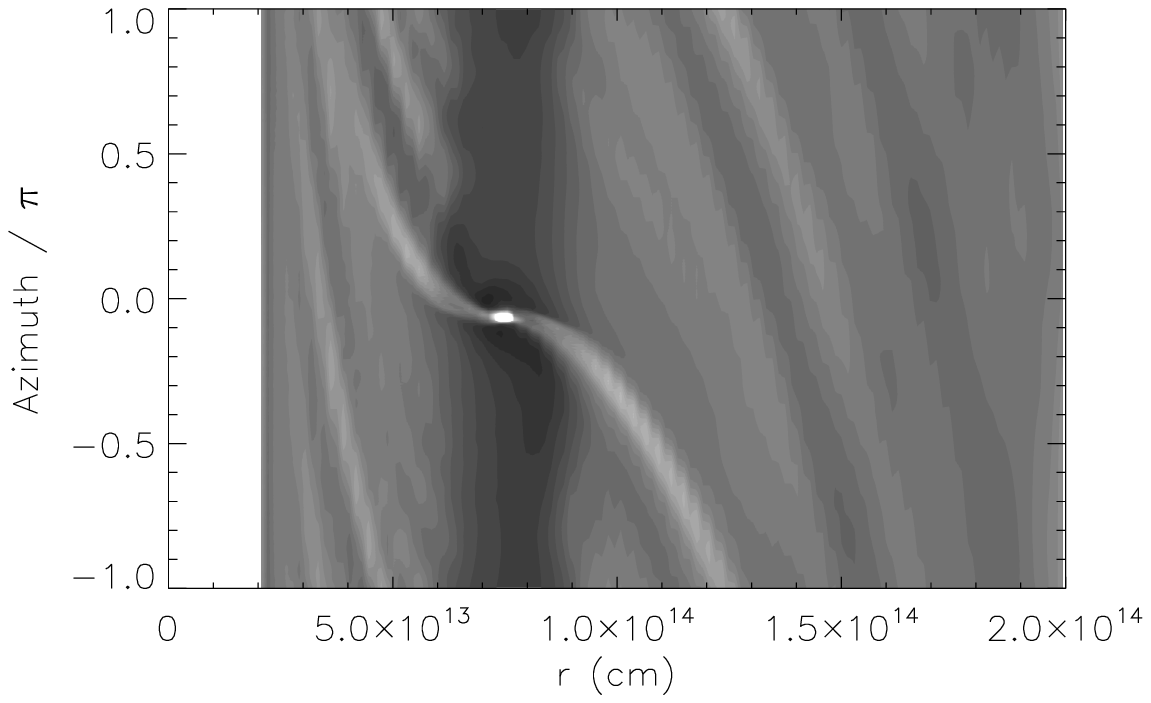}}
\includegraphics[scale=0.6]{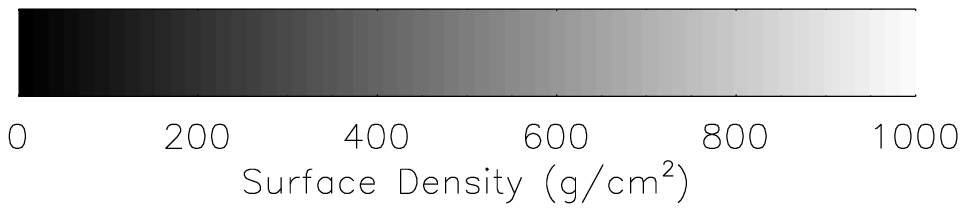}
\caption{Surface density after 100 orbits for a $q=10^{-3}$ planet in the zero (top), low (middle) and high (bottom) viscosity cases. The scale is linear, and ranges between 0 and \unit{1000}{\grampersqcmnp}. Note that since we only model the disc for $z>0$, the unperturbed surface density is \unit{500}{\grampersqcmnp}}
\label{fig:JupiterSurfaceDensity}
\end{figure}

We will now discuss the results we obtained from a Jupiter mass ($q=10^{-3}$) planet embedded in the zero, low and high viscosity discs.
In Figure~\ref{fig:JupiterSurfaceDensity}, we show the surface density of the disc after one hundred orbits.
Note that although the total surface density of our disc is \unit{1000}{\grampersqcmnp}, we are only computing the portion of the disc with $z > 0$.
Hence the initial surface density was only \unit{500}{\grampersqcmnp} in the computational volume.
The behaviour we see is expected \citep[see, e.g. the comparisons of][]{2006MNRAS.370..529D}.
The zero viscosity case has lots of fine structure and a very clean gap.
The low viscosity case has a clean gap, but most of the fine structure has been suppressed, while the gap in the high viscosity case has been suppressed (although we can still see a substantial depression in the surface density).

\begin{figure*}
\centering
\subfigure{\includegraphics[scale=0.6]{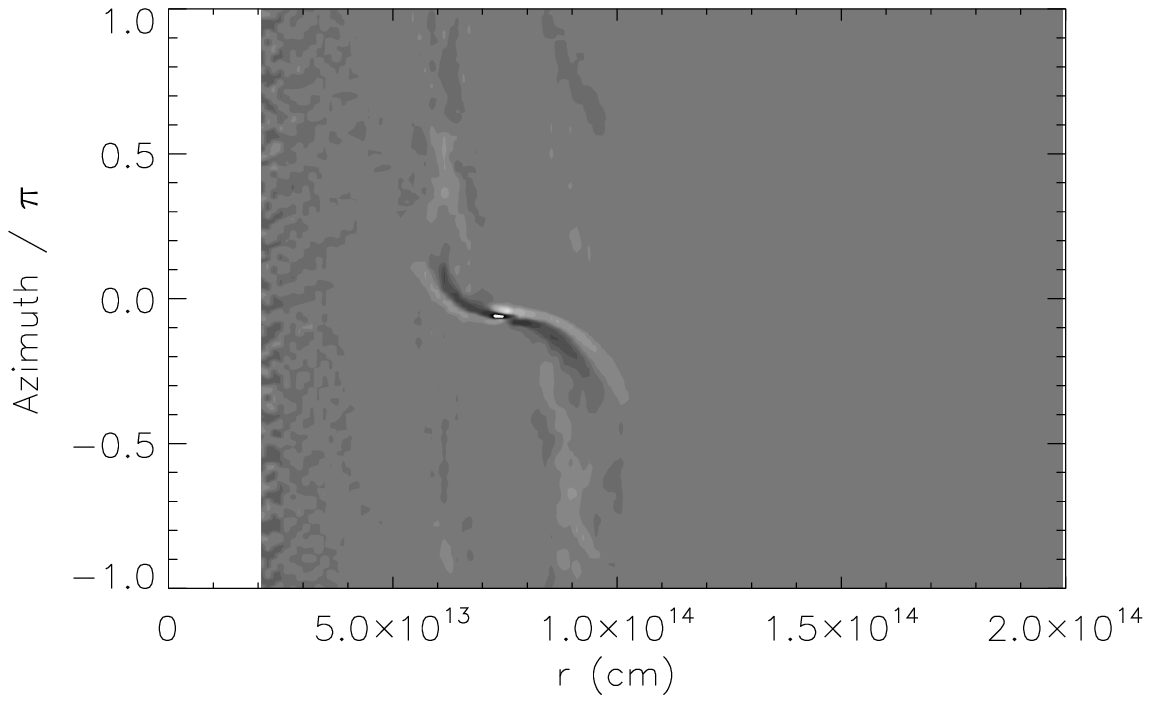}
           \includegraphics[scale=0.6]{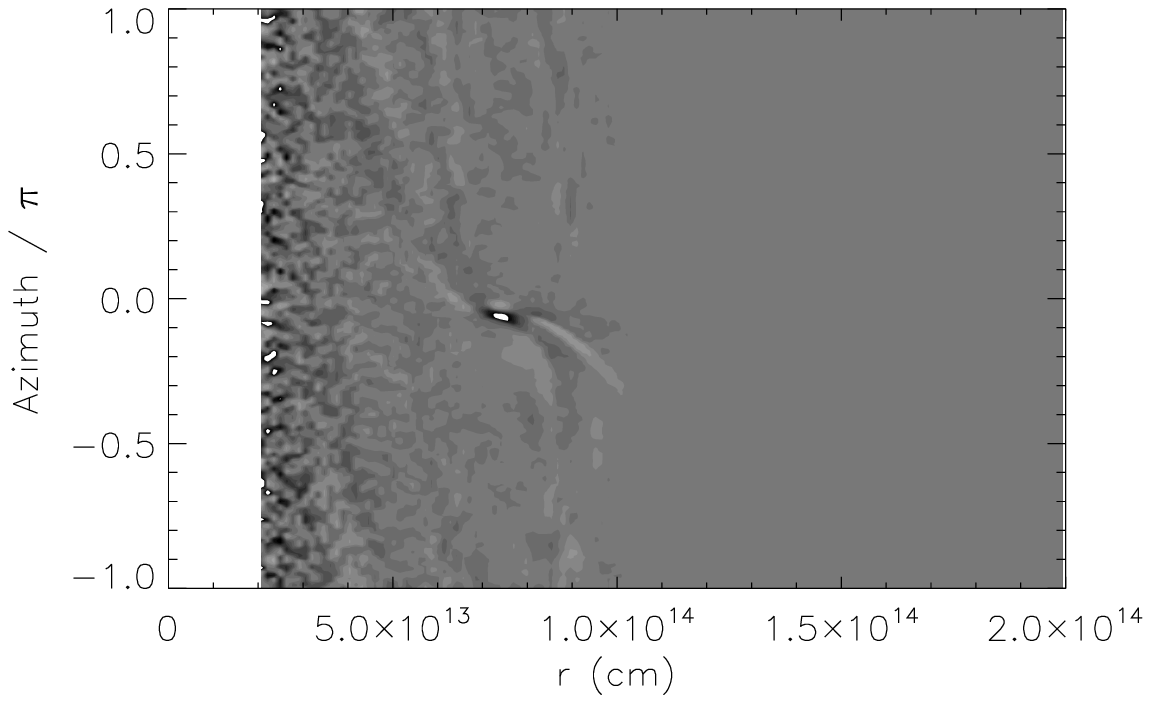}}
\subfigure{\includegraphics[scale=0.6]{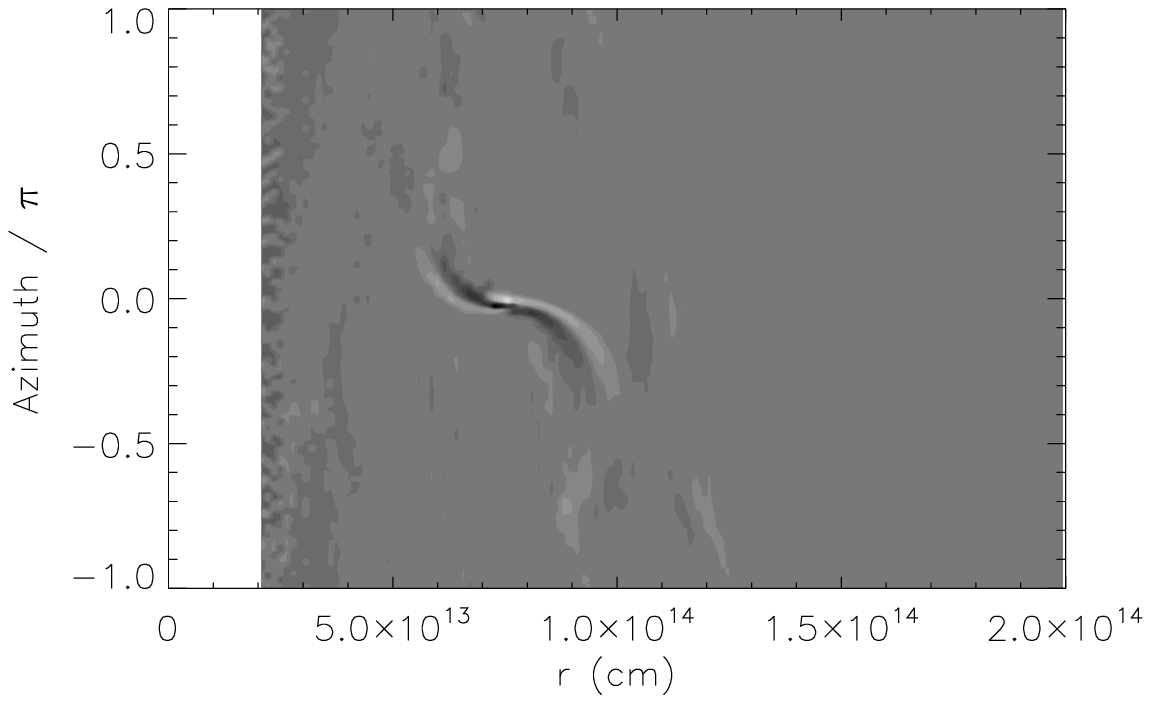}
           \includegraphics[scale=0.6]{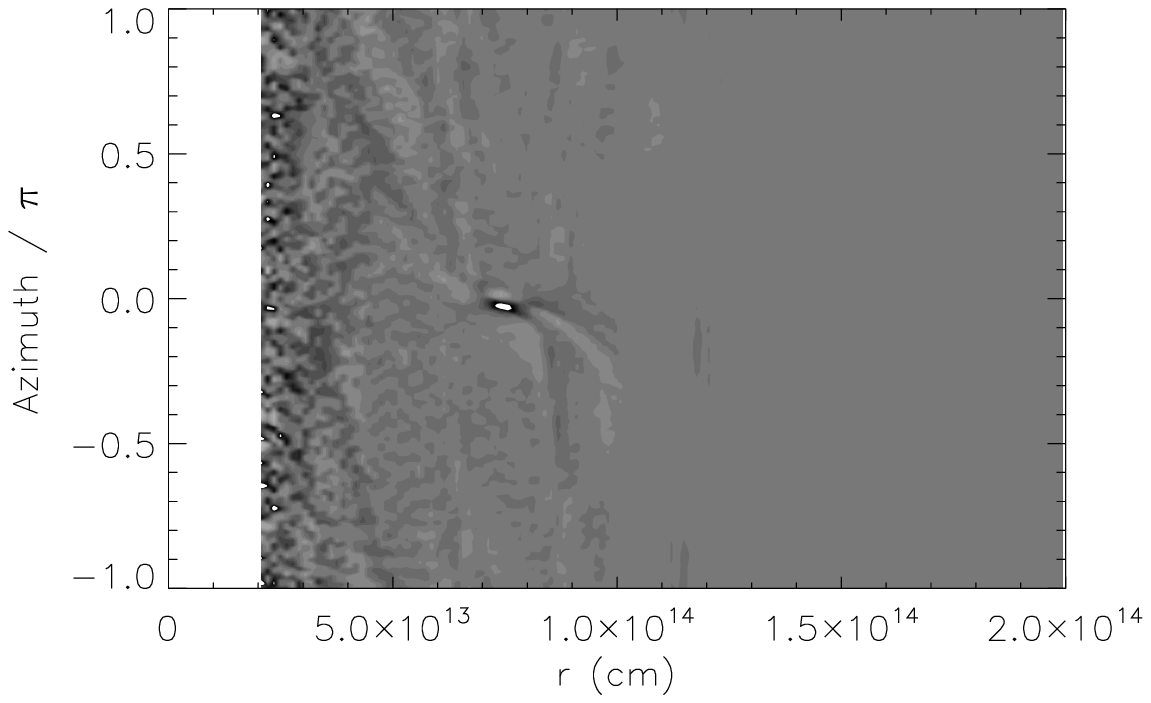}}
\subfigure{\includegraphics[scale=0.6]{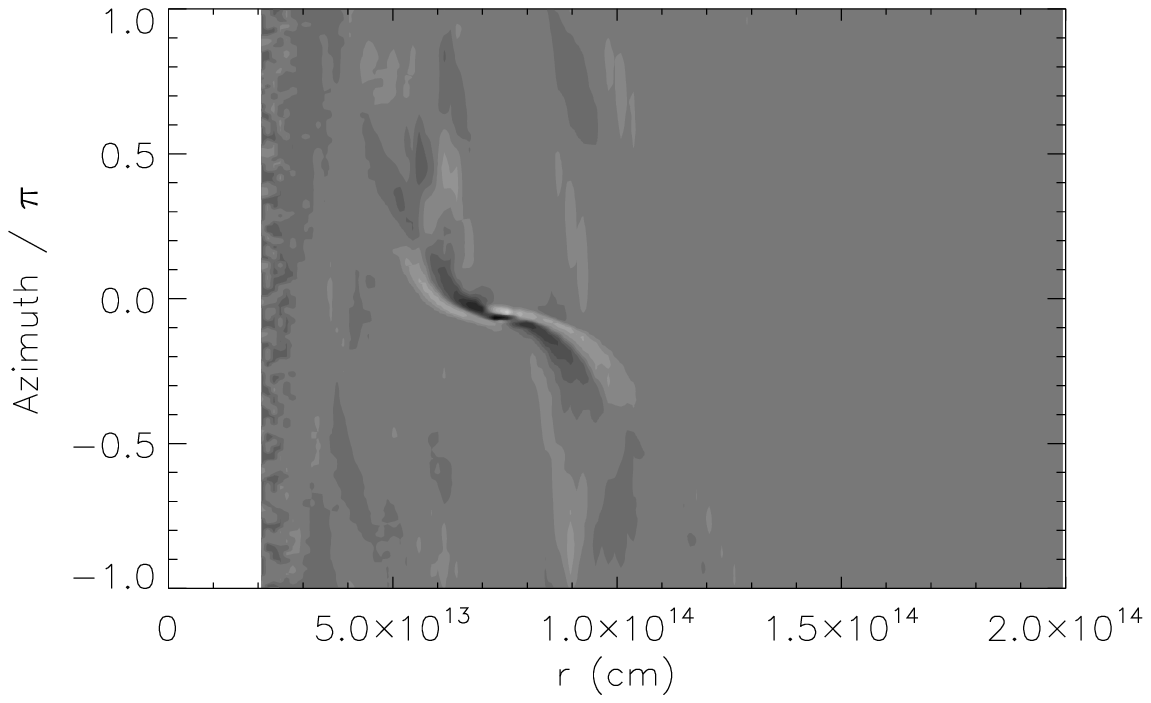}
           \includegraphics[scale=0.6]{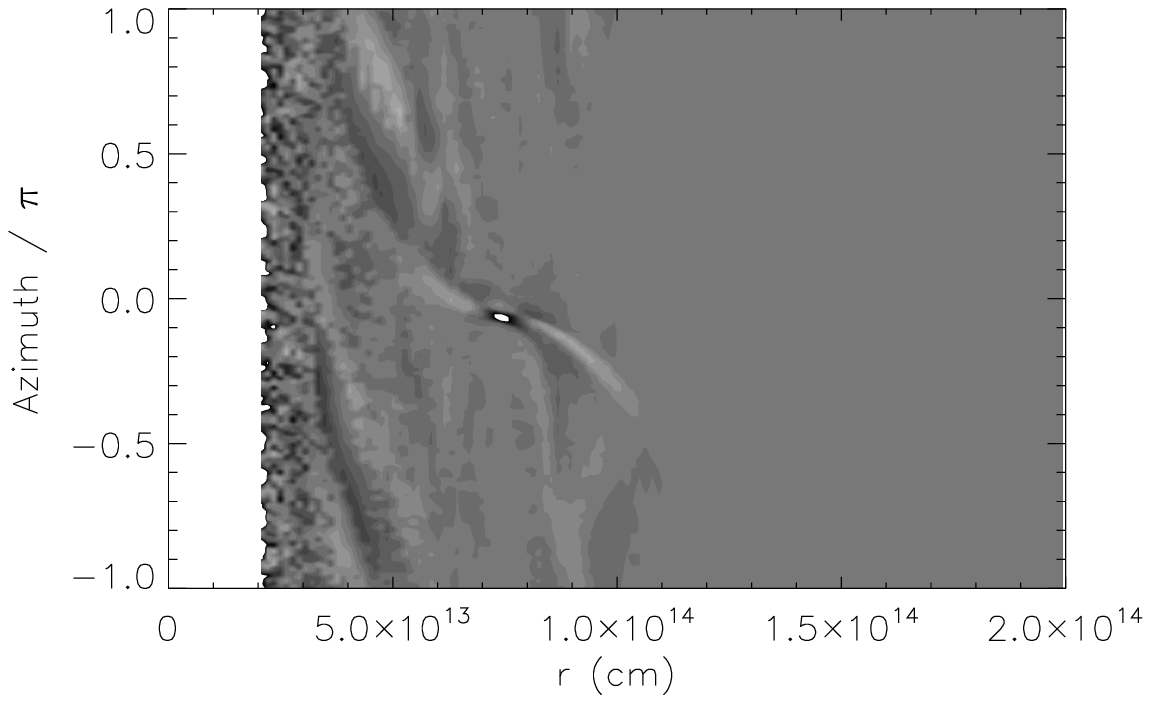}}
\includegraphics{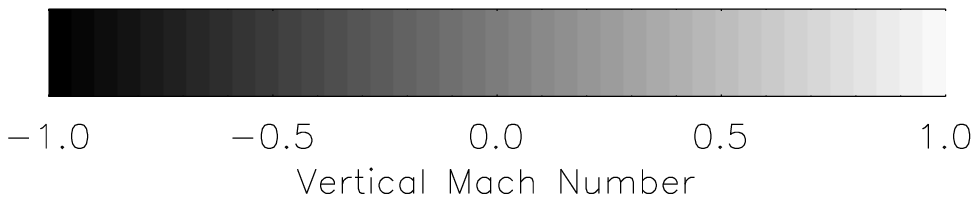}
\caption{Mach number of the vertical velocity on planes with $z=\rHill{}/4$ (left) and $\rHill{}$ (right) for a $q=10^{-3}$ planet in the zero (top), low (middle) and high (bottom) viscosity cases. These are shapshots taken after 100 orbits. The scale ranges from $\Machz{}=-1$ (black) to 1 (white)}
\label{fig:MachZSliceZJupiter}
\end{figure*}

In Figure~\ref{fig:MachZSliceZJupiter} we plot the vertical Mach number, $\Machz{} \equiv v_z / c_s$ in the disc for slices with $z=\rHill{}/4$ and $z=\rHill{}$.
Unsurprisingly, there are strong motions close to the planet.
However, we also see vertical motions associated with the spiral arms seen in Figure~\ref{fig:JupiterSurfaceDensity}.
These motions are relatively strong (up to $\Machz{} \approx 0.5$), and reach well into the gap edge (although they do not propagate as far as the spiral arms in the density field..
The strength of these motions does not appear to be affected by the viscosity, with a similar structure appearing for the zero, low and high viscosity cases.
This is not especially surprising, since viscous effects take several orbits to become apparent, whereas the planet stirs each part of the disc every orbit.
Furthermore, we only apply the \mbox{$r$--$\phi$} component of the viscous stress tensor, so the vertical motion can only be affected indirectly.

Averaging the motion around an orbit, we find that the mean velocity is zero, indicating that the vertical motions are waves.
Given their location, we conclude that these vertical motions are the \textsc{3d} aspect of the spiral arms generated by the planet.
Previous studies \citep[see][ and references therein]{2000prpl.conf.1135W} of the spiral arms generated by a planet have usually assumed that the disc is two dimensional.
Here, we see that the spiral arms have a definite three dimensional structure.
The gas contracts and expands vertically as it passes through the spiral arms.
However, this \textsc{3d} structure is only apparent close to be planet, as the vertical motions rapidly fade at larger radial distances.
This can be understood in terms of the thin disc approximation and the planet's Hill sphere.
If a disc is thin, then it may be treated as a \textsc{2d} entity \citep{1981ARA&A..19..137P}.
The discs used in this paper have $h/r$ set according to Equation~\ref{eq:flaringsetup}, and as such are relatively thin.
Accordingly, we should expect the waves to be \textsc{2d} in nature.
However, within the planet's Hill sphere the planet's gravity dominates.
This has a strong vertical component, which drives the \textsc{3d} structures we see in Figure~\ref{fig:MachZSliceZJupiter}.
As these waves propagate beyond the planet's Hill sphere, the thin disc approximation reasserts itself, and the vertical motions dissipate.

There are a number of possible numerical issues with our results.
Firstly, there is our choice of equation of state.
We imposed a temperature structure on the disc, but used $\gamma=1.1$ in the Riemann solver.
Of course, using an isothermal Riemann solver would not be fully self-consistent either, since there is a radial temperature gradient.
In Appendix~\ref{sec:changegamma}, we demonstrate that reducing $\gamma$ to a value of $1.01$ (thereby making the Riemann solver more closely approximate the isothermal case) does not have a significant effect on our results.

The viscosity is also an issue.
We chose to include only the \mbox{$r$--$\phi$} component of the viscous stress tensor, which is sufficient to drive accretion.
The true source of angular momentum transport in astrophysical discs is unlikely to act like a simple physical viscosity, so including components not essential to accretion did not seem necessary.
This said, it is important to verify that this does not produce any important inconsistences.
In Appendix~\ref{sec:viscouscomponents}, we show that including all components of the viscous stress tensor in our runs does not affect the vertical oscillations we observe.

Finally, our cylindrical grid has relatively poor resolution (compared to $h$) in the innermost regions of the disc, so Figure~\ref{fig:MachZSliceZJupiter} is somewhat noisy there.
However, the spiral arms remain clearly visible, indicating that the noise is not dominating our results.
Nevertheless, we performed a short resolution test, detailed in Appendix~\ref{sec:resolution}.
We show that doubling the resolution supresses the noise seen in Figure~\ref{fig:MachZSliceZJupiter}, with little effect on the vertical oscillations induced by the planet.


\subsection{Two Jupiter Mass Planet}

In this section, we shall discuss the results obtained when a \unit{2}{\Mjup} planet orbits in the same three discs.
For this case, we found that the buffer zone damping time \citep[see][]{2006MNRAS.370..529D} had to be reduced to half an orbit, to suppress wave reflection.

\begin{figure}
\centering
\subfigure{\includegraphics[scale=0.6]{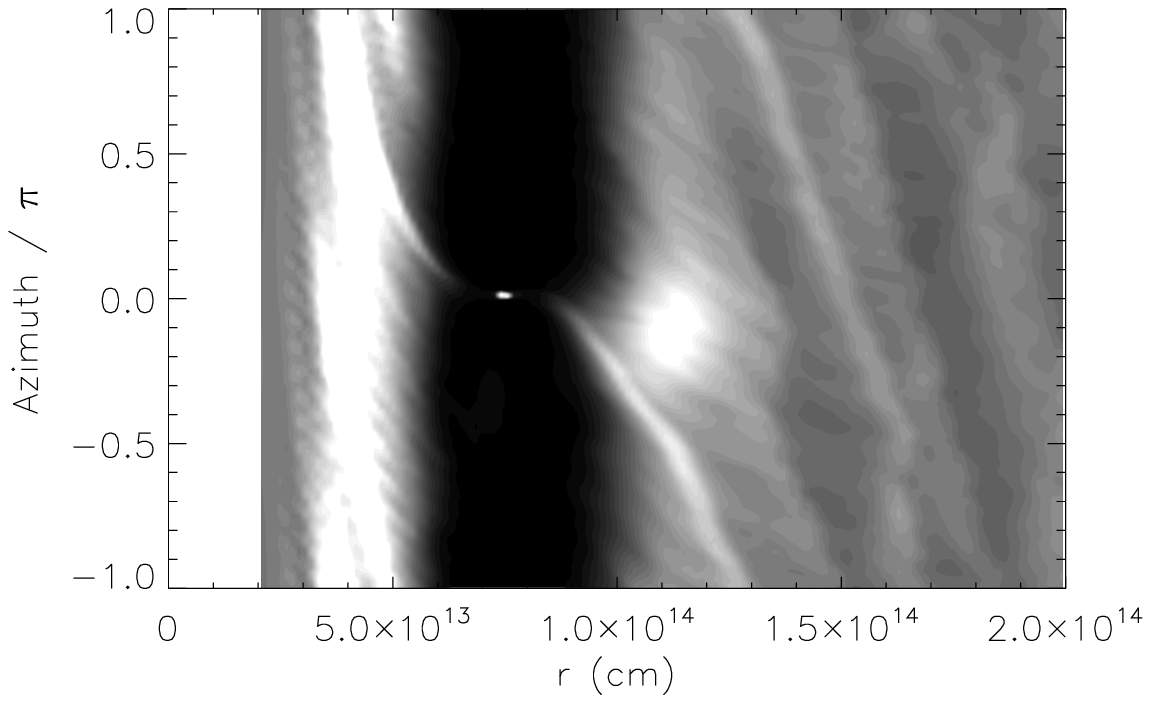}}
\subfigure{\includegraphics[scale=0.6]{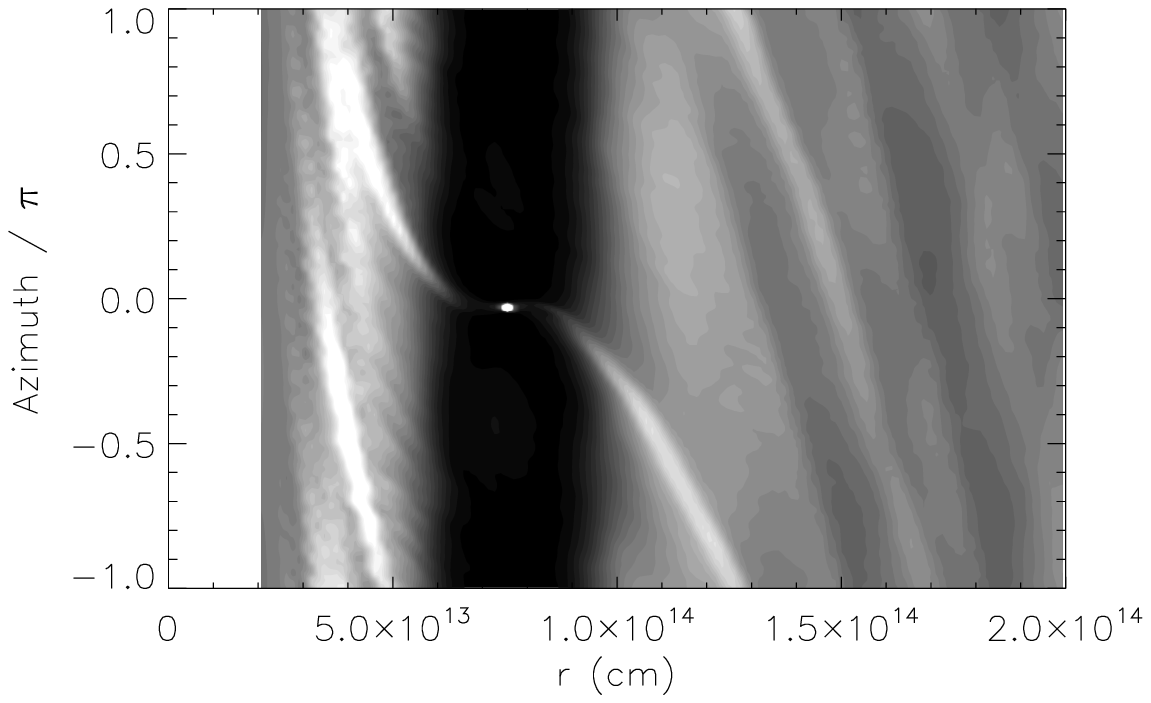}}
\subfigure{\includegraphics[scale=0.6]{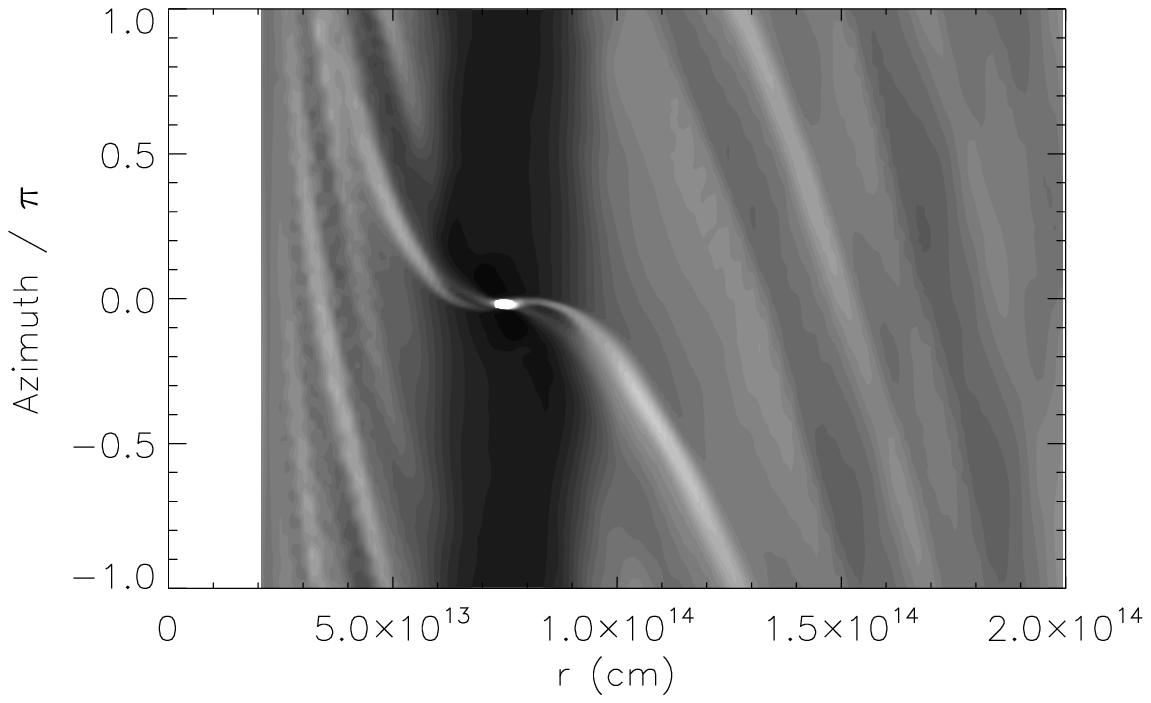}}
\includegraphics[scale=0.6]{SigmaBar}
\caption{Surface density after 100 orbits for a $q=\scinot{2}{-3}$ planet in the zero (top), low (middle) and high (bottom) viscosity cases. The scale is linear, and ranges between 0 and \unit{1000}{\grampersqcmnp}. Note that since we only model the disc for $z>0$, the unperturbed surface density is \unit{500}{\grampersqcmnp}}
\label{fig:TwoJupiterSurfaceDensity}
\end{figure}

We start with Figure~\ref{fig:TwoJupiterSurfaceDensity}, which shows the surface density in the disc after approximately 100 orbits of the planet.
This can be compared to Figure~\ref{fig:JupiterSurfaceDensity}.
We see that a substantial gap has appeared in all cases, although there is some material visible in the gap for the high viscosity case.
In the zero viscosity case, we can see a density enhancement moving along the gap edge.
When viscosity is present, it appears that formation of this enhancement is suppressed.
As in the case of Figure~\ref{fig:JupiterSurfaceDensity}, the surface density plots are much as expected.

\begin{figure*}
\centering
\subfigure{\includegraphics[scale=0.6]{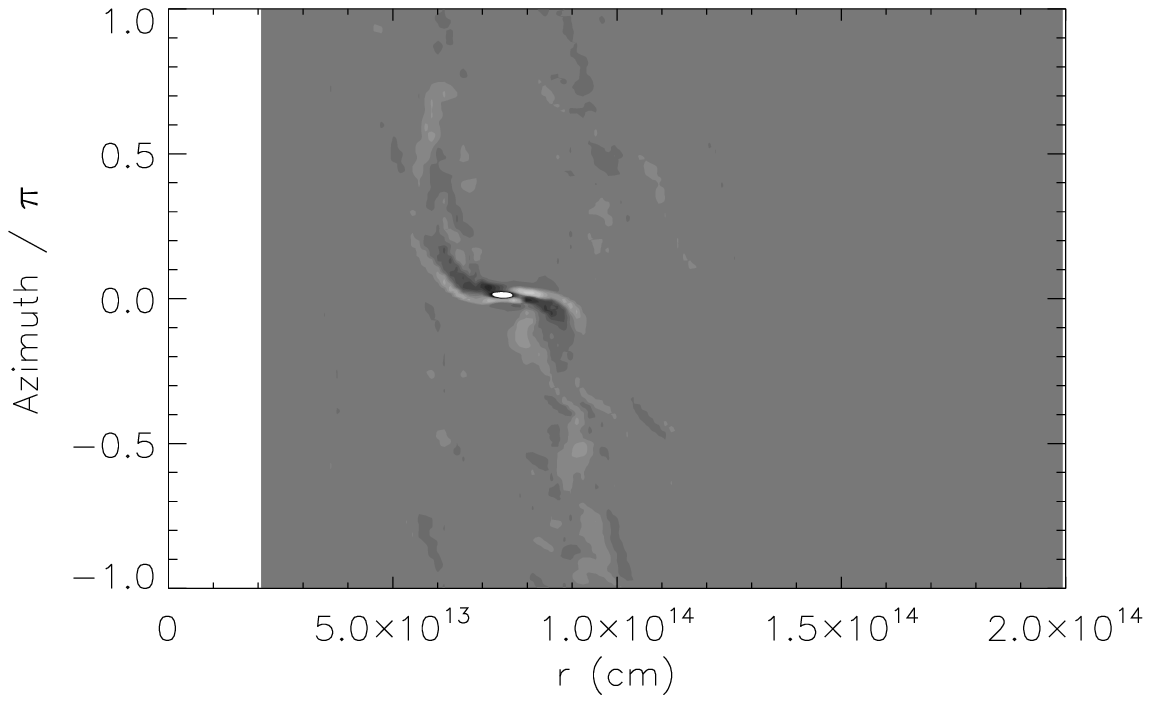}
           \includegraphics[scale=0.6]{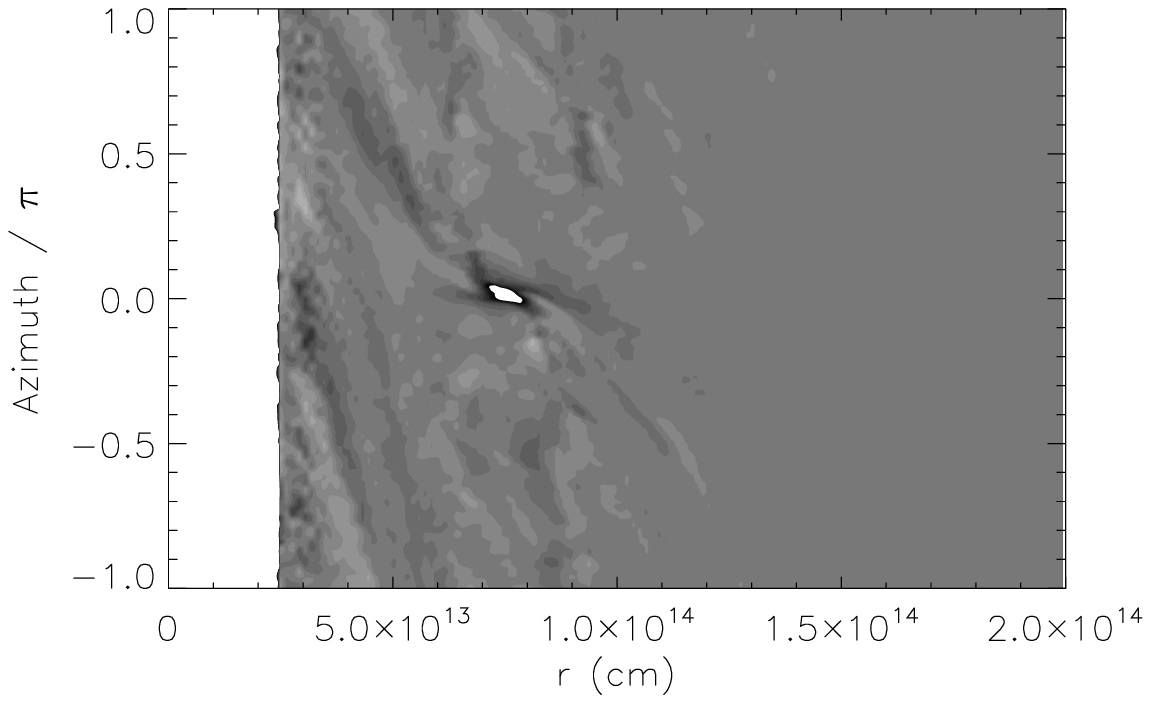}}
\subfigure{\includegraphics[scale=0.6]{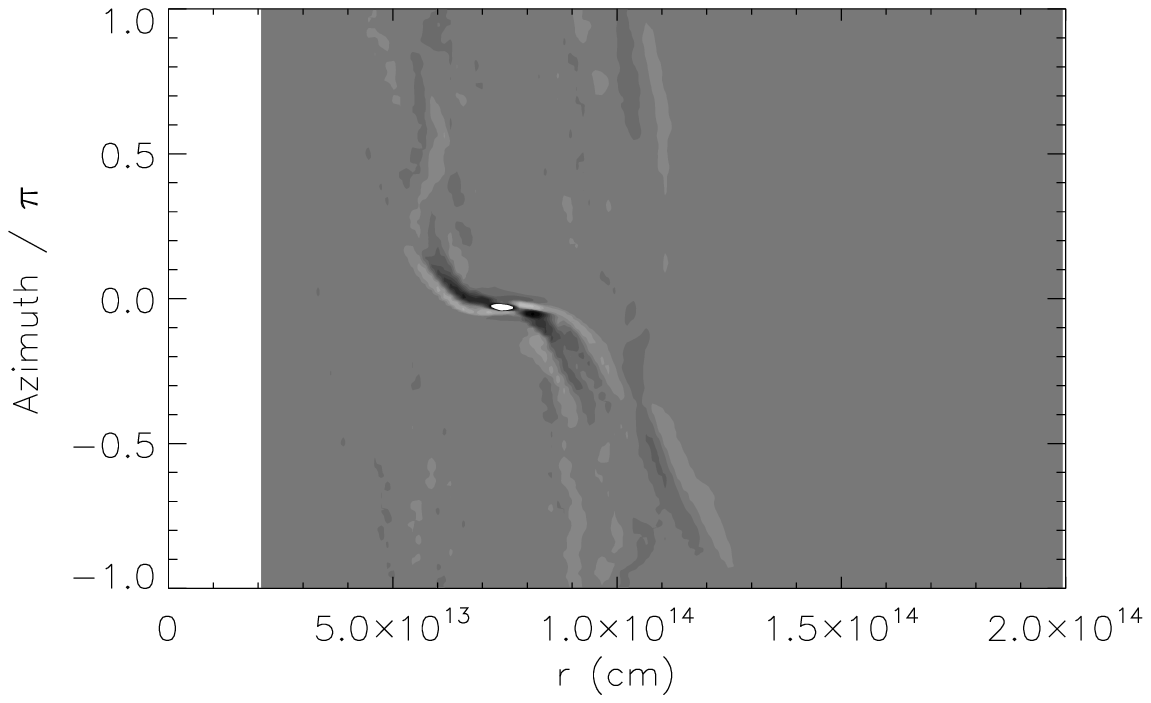}
           \includegraphics[scale=0.6]{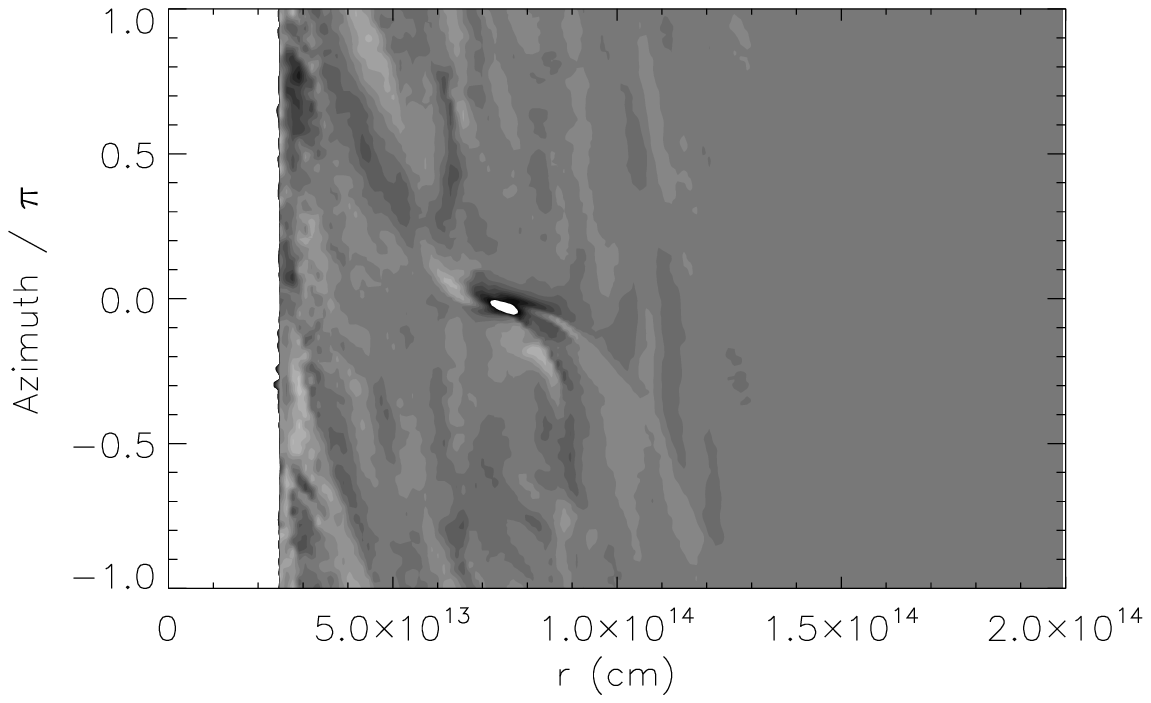}}
\subfigure{\includegraphics[scale=0.6]{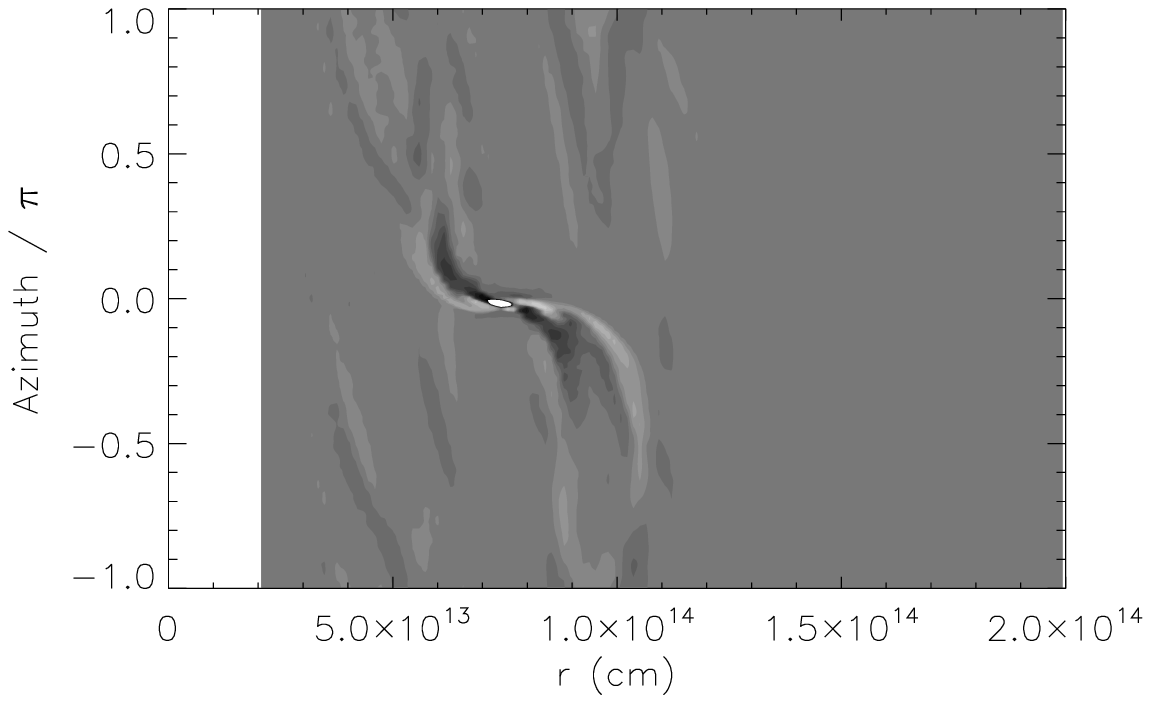}
           \includegraphics[scale=0.6]{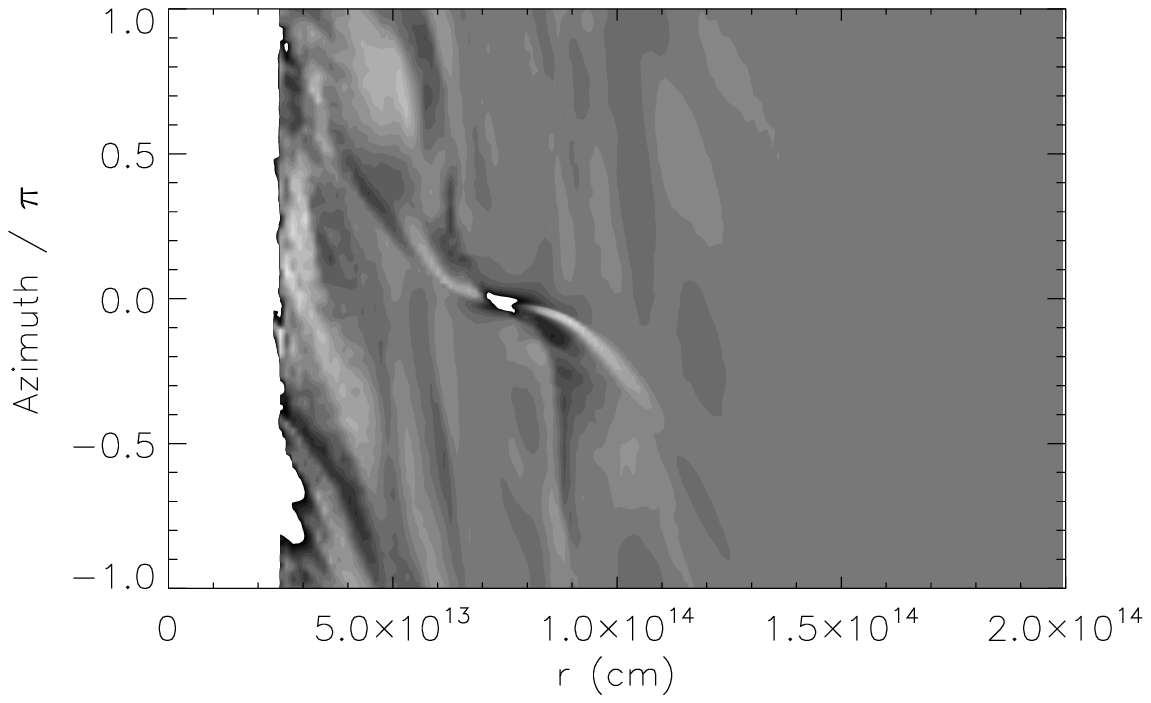}}
\caption{Mach number of the vertical velocity on planes with $z=\rHill{}/4$ (left) and $\rHill{}$ (right) for a $q=\scinot{2}{-3}$ planet in the zero (top), low (middle) and high (bottom) viscosity cases. The discs are shown after 100 orbits of the planet. The scale ranges from $\Machz{}=-1$ (black) to 1 (white)}
\label{fig:MachZSliceZTwoJupiter}
\end{figure*}

Figure~\ref{fig:MachZSliceZTwoJupiter} shows \Machz{} on planes with $z=\rHill{}/4$ and $z=\rHill{}$, and may be directly compared to Figure~\ref{fig:MachZSliceZJupiter}.
The \Machz{} arms are more prominent, particularly on the $z=\rHill{}$ surface.
However, the velocities reached are not substantially greater and the radial propagation of the velocity field does not appear to be greater.
We conclude that increasing the planet's mass increases the vertical domain over which the planet can drive vertical motions, but does not substantially increase their maximum strength or radial domain.

The vertical domain of the vertical oscillations expands because the driving is stronger.
Doubling the mass of the planet will double its gravitational field, increasing the vertical distance over which driving is effective.
However, the extent of the driving in the orbital plane is still limited by the size of the Hill sphere.
Since $\rHill{} \propto q^{1/3}$, we should not expect a \unit{2}{\Mjup} planet to excite vertical oscillations over a substantially larger radial domain than a \unit{1}{\Mjup} planet.


\subsection{Low Mass Planet}

In addition to the two sets of runs described above, we performed a set of runs with a \unit{0.1}{\Mjup} planet.
As predicted, this planet did not open a gap, but merely caused a depression in the surface density.
Furthermore, we did not find an substantial vertical motions being driven, in contrast to those shown in Figures~\ref{fig:MachZSliceZJupiter} and~\ref{fig:MachZSliceZTwoJupiter}.
This is consistent with the arguments about the planet's Hill sphere presented above.
The Hill sphere of a \unit{0.1}{\Mjup} planet fits entirely within one vertical scale height.
Since the planet's gravity only dominates within the Hill sphere, it is then not surprising that the vertical component of the planet's gravity is unable to drive strong vertical oscillations.
The region it affects is simply too small.

%% file: discuss.tex
\section{Discussion}
\label{sec:discuss}

In Section~\ref{sec:results} we presented an analysis of the vertical velocity structure induced by a Jupiter-mass planet embedded in a circumstellar disc.
Significant vertical motions were induced in the spiral arms raised by the planet, with vertical Mach numbers up to $\approx 0.5$.
We shall now discuss the possible observable consequences.

\subsection{Thickness of the Gap Edge}

Vertical motions induced by the spiral arms have the potential to alter the appearance of the gap edge in a number of ways.
Since the gap edge is likely to be the first feature which will be imaged directly, it is important to understand these.

Firstly, if the inner disc is depleted (not in our current calculations, but recall systems such as CoKu Tau/4, DM Tau and GM Aur), then stellar radiation will directly strike the gap edge, and heat it.
This would increase the scale height, but only modestly since $h \propto c_s \propto \sqrt{T}$.
The edge of the gap would also appear to be very bright, due to its large directly illuminated surface area.
A more significant increase may come from the density variations in the spiral arms themselves.
Across the spiral arm, densities vary by a factor of a few.
However, since the density distribution in a vertically isothermal disc is a gaussian, the scale height only depends on the log of this.
We might expect the spiral arms (if not resolved) to increase the apparent scale height by $\approx h$.
We note that \citet{2002ApJ...566L..97W} made simulated observations of a similar situation, and found that the spiral arms were difficult to see.
However, their work appears to have been based on a \textsc{2d} hydrodynamic code, which would suppress the strong vertical motions we found.

Since instruments such as \emph{Spitzer} observe the dust (and not the gas), it is important to consider the effect of the coupling between the gas and dust.
The strong vertical motions imply that gas streamlines move up and down about half the scale height.
Dust coupled to gas would also move up and down by the same distance.
The strong vertical streaming motions imply that there are larger velocity gradients present than expected from a stratified or sedimented disk.
We expect these streaming motions to suppress dust sedimentation, as compared with a standard turbulent disc.

\citet{2004A&A...421.1075D} developed a simple model for dust sedimentation, and applied it to evolving protoplanetary discs.
In their models, dust grains larger than around \unit{1}{\micro\metre} were sufficiently decoupled from the gas to undergo significant sedimentation, potentially affecting the expected spectral energy distribution (SED) obtained from the discs.

In their model, \citeauthor{2004A&A...421.1075D} balanced the stirring and settling timescale of the dust.
Settling was due to the vertical component of gravity within the disc.
Balancing this against Epstein drag (which should be applicable to all dust particles within a protoplanetary disc), they determined a settling time of
\begin{equation}
\tset{} = \frac{4}{3 \sqrt{2 \pi}} \frac{\sigma}{m} \frac{\Sigma(r)}{\OmegaKep{}} \exp \paren{- \frac{z^2}{2 h^2(r)}}
\label{eq:tsettle}
\end{equation}
where the grains have mass $m$, and cross section $\sigma$.
Stirring was provided by the viscous processes in the disc.
Particles would diffuse upwards, under the influence of eddies in the disc, so the stirring timescale would occur on a timescale $\tstir{} = z^2 / D$, where $D$ is the effective diffusion coefficient.
Following \citeauthor{2004A&A...421.1075D}:
\begin{equation}
\tstir{} = \frac{\text{Sc}}{\alpha \OmegaKep{}} \frac{z^2}{h^2(r)}
\label{eq:tstir}
\end{equation}
where we have assumed that we have a standard $\alpha$-disc, with  $\nu = \alpha c_s h$, and also that $D \equiv \nu$.
Where $\text{Sc}$ is the Schmidt number of the dust grains in the gas, which measures the strength of the coupling between the gas and the dust.
In our discs, $\text{Sc}$ only becomes significantly greater than $1$ around five scale heights above the midplane.
Accordingly, we shall take $\text{Sc}=1$ in the current discussion.
See section~2 of \citet{2004A&A...421.1075D} for detailed exposition of dust settling.
Although we used a constant viscosity disc, we can define an equivalent $\alpha$ value, finding it to be $\alpha \approx \scinot{2.9}{-3}$ at the planet's orbital radius for the low viscosity disc.
The settling height is then defined by requiring $\tstir{} = \xi \tset{}$, with $\xi$ typically one hundred.

Our numerical experiments make use of a much denser gas disc than \citeauthor{2004A&A...421.1075D} -- see particularly their equation~19.
Their disc is depleted by a factor of roughly 30, as compared with ours.
This strengthens the coupling between the dust and gas, in turn increasing the time required for dust to settle to the midplane.
\citet{2004A&A...425L...9P} made use of a disc depleted by a similar amount, in their work showing decoupling between the gas and dust discs.
They also followed grains of size \unit{1}{\milli\metre}, which are observable to instruments such as the \emph{SMA} and \emph{ALMA}, but less easily detectable by \emph{Spitzer}.
We would not expect our disc to sediment at all.
However, since we do not include self-gravity and impose a temperature profile, the densities in our numerical experiments can easily be rescaled.

At later times, the disc would be depleted and we would expect sedimentation along the lines of the theory developed by \citeauthor{2004A&A...421.1075D}.
However, we must also consider the stirring provided by the planet.
As shown in Section~\ref{sec:results}, there are strong vertical motions in the spiral arms within the gap edge.
The vertical stirring caused by the spiral arms occurs on the synodic timescale, over approximately one scale height.
The synodic timescale is given by $\Omega_{\text{syn}} \equiv \Omega - \OmegaPlanet{}$, and varies with position in the disc.
However, we are interested in particles close to the gap edge, which invariably lies close to the $m=2$ Lindblad resonances.
The synodic period will therefore be about twice the orbital period.
A particle  that is lifted upwared by spiral arm and then diffuses out of the arm would enter a stream line with a greater mean height. 
This could raise the effective diffusion coefficient for the dust grains from $D \approx \alpha c_s h$ to $D \approx 0.5 c_s h$.
Comparing to our equivalent $\alpha = \scinot{2.9}{-3}$, we see that the stirring timescale implied by Equation~\ref{eq:tstir} would fall by over two orders of magnitude.
However, the settling time (Equation~\ref{eq:tsettle}) is unaffected by the spiral arms, and so will remain the same.
Because the ratio of the settling timescale to the stirring timescale depends on the local gas density (which roughly follows a gaussian distribution vertically), we expect the depletion height to depend on the square root of the log of the diffusion coefficient.
Since our effective diffusion coefficient increases by a factor of $\approx \alpha$, we expect the depletion height to be raised by a few scale heights.

In summary, three different effects lead to apparent thickening of edges of discs near planets
\begin{enumerate}
\item Direct stellar radiation raising the scale height and temperature in the gap edge.
Size of effect: Factor of a few increases in temperature  but scale height only depends on square root of temperature so increase in scale height is only modest

\item Density variations caused by spiral arms lead to raising of $\tau=1$ surface.
Density contrasts in spiral arms are a factor of a few but variation in scale height depends on the natural log of this.
Size of effect $\sim 1$ scale heights

\item Increased vertical mixing of dust caused by the vertical motions in the spiral arms.
The  stirring timescale is effectively increased by $\alpha^{-1}$.
The depletion height could vary by the square root of the log of this, which is a few scale heights
\end{enumerate}
Combined, these effects can make the inner edge of the gap appear significantly thicker than the bulk of the disc.
We note that \cite{2005ApJ...621..461D} calculated an aspect ratio of $h/r \approx 0.1$ for the edge of CoKuTau/4's disc, which seems quite large for a disc \unit{11}{\AU} from a \unit{0.5}{\Msol} star.
In addition to these effects, the vertical motions themselves will make some portion of the disc edge appear higher - perhaps by as much as half a scale height.
However, since the vertical motions are waves, other portions of the disc edge will be depressed by a similar amount.
Whether this extra thickening (or thinning) is observable will therefore depend on the viewing angle and orbital phase of the planet, especially in azimuthally averaged spectral energy distributions.

In a fully self-consistent model, with radiative transfer and dust settling, the increased scale height might cause the gap to become shallower, as the tidal condition introduced at the start of Section~\ref{sec:results} fails.
It is also possible that the thick disc edge will cast a shadow over the outer disc, causing it to cool and become geometrically thinner \citep[see][ for a discussion of self-shadowing in the context of Herbig Ae/Be stars]{2004A&A...417..159D}.
The light observed from such a disc would be completely dominated by the bright inner edge of the gap induced by the planet.

\subsection{Detection of the Velocity Field}

Almost all extra-solar planets have been discovered by the radial velocity (RV) method.
Although its success is unquestioned, the RV method suffers from two strong biases.
Firstly, it is most sensitive to massive planets in close orbits.
Secondly, it can only be used on old stars ($t > \unit{1}{\giga\yyear}$), or surface activity will wipe out the small signal from a companion.
We would greatly deepen our understanding of planet formation, if we could catch planet formation in progress.
New instruments, primarily \emph{ALMA}, offer this possibility.
\citet{2002ApJ...566L..97W,2005ApJ...619.1114W} have already argued that the `accretion hotspot' of an embedded protoplanet will be detectable in \emph{ALMA} images.
We shall now discuss the prospects of using \emph{ALMA}'s high velocity resolution to detect the spiral arms raised by a protoplanet.

\citet{2003A&A...399..773D} examined the structure of DM Tau's outer ($r > \unit{50}{\AU}$) accretion disc in several CO lines.
They found that the velocity dispersion in the disc midplane was higher than that towards the surface.
This implies that the turbulence is greater in the midplane, even though they found that the disc surface was warmer than the midplane.
A natural explanation for this would be spiral arms stirring up the midplane.
A planet is unlikely so far out in the disc, but gravitational instability would be quite likely.

Even with its \unit{0.05}{\arcsecond} resolution, the beam of \emph{ALMA} will still be approximately \unit{5}{\AU} across at the nearest star-forming clouds (located approximately \unit{100}{\parsec} away from us).
This means that any effects of a planet in a \unit{5}{\AU} orbit are highly unlikely to be visible.
Accordingly, we have rescaled our results to place the planet on a \unit{10}{\AU} orbit.
We computed the line of sight velocity for every grid cell in our computational domain, and projected this onto the sky.
We then traced rays through the projected density structure, evalutating the integral
\begin{equation}
\Sigma_{\text{los}} = \int_{\infty}^{l} \rho \sdiffer{l}
\end{equation}
We assumed that \emph{ALMA} would observe the velocity in the grid cell where $\Sigma_{\text{los}} = 1$.
Although this isn't necessarily the surface \emph{ALMA} would see, it serves as a useful initial estimate.

\begin{figure}
\centering
\begin{tabular}{cc}
\includegraphics[width=4cm]{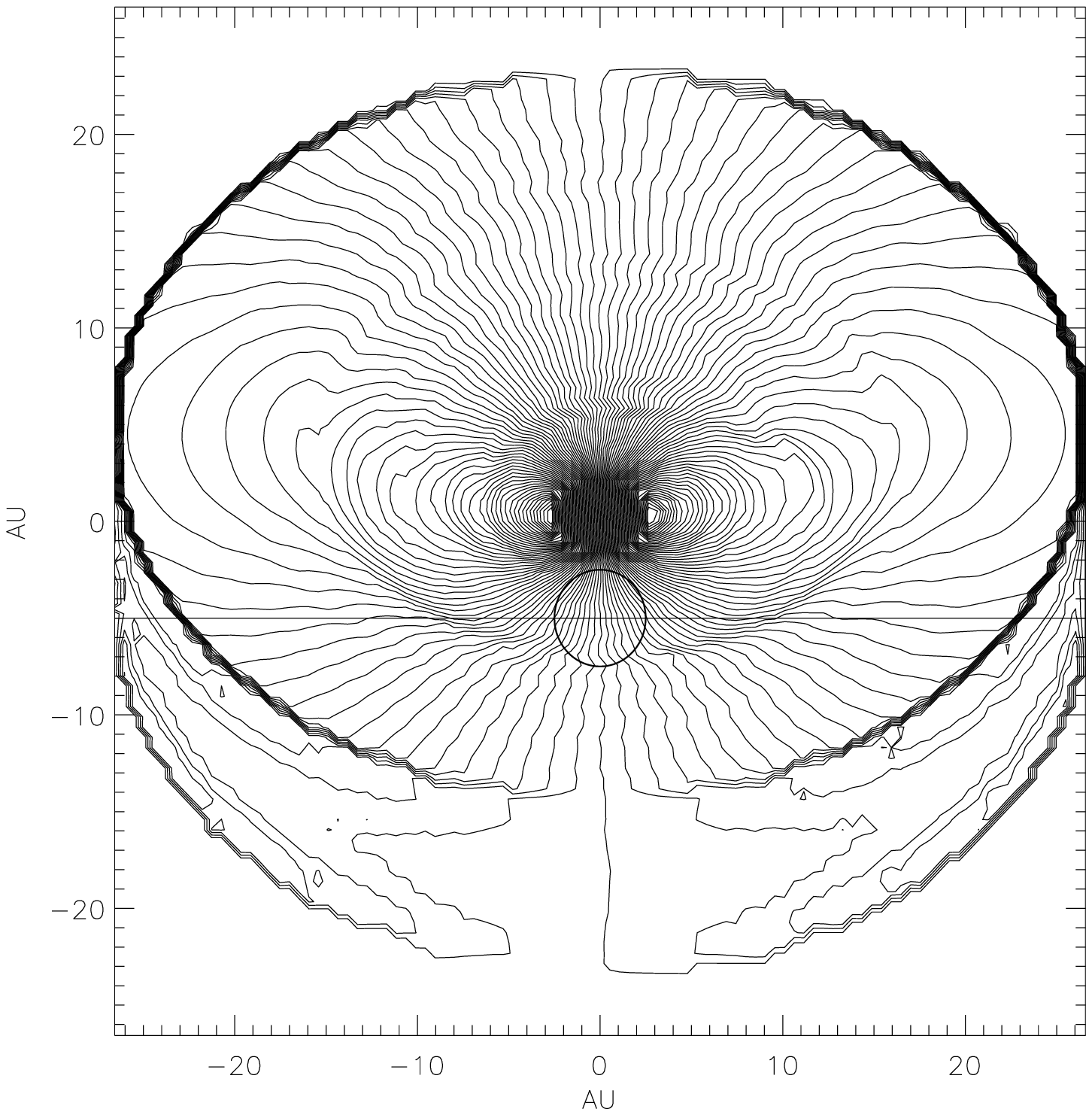} &
\includegraphics[width=4cm]{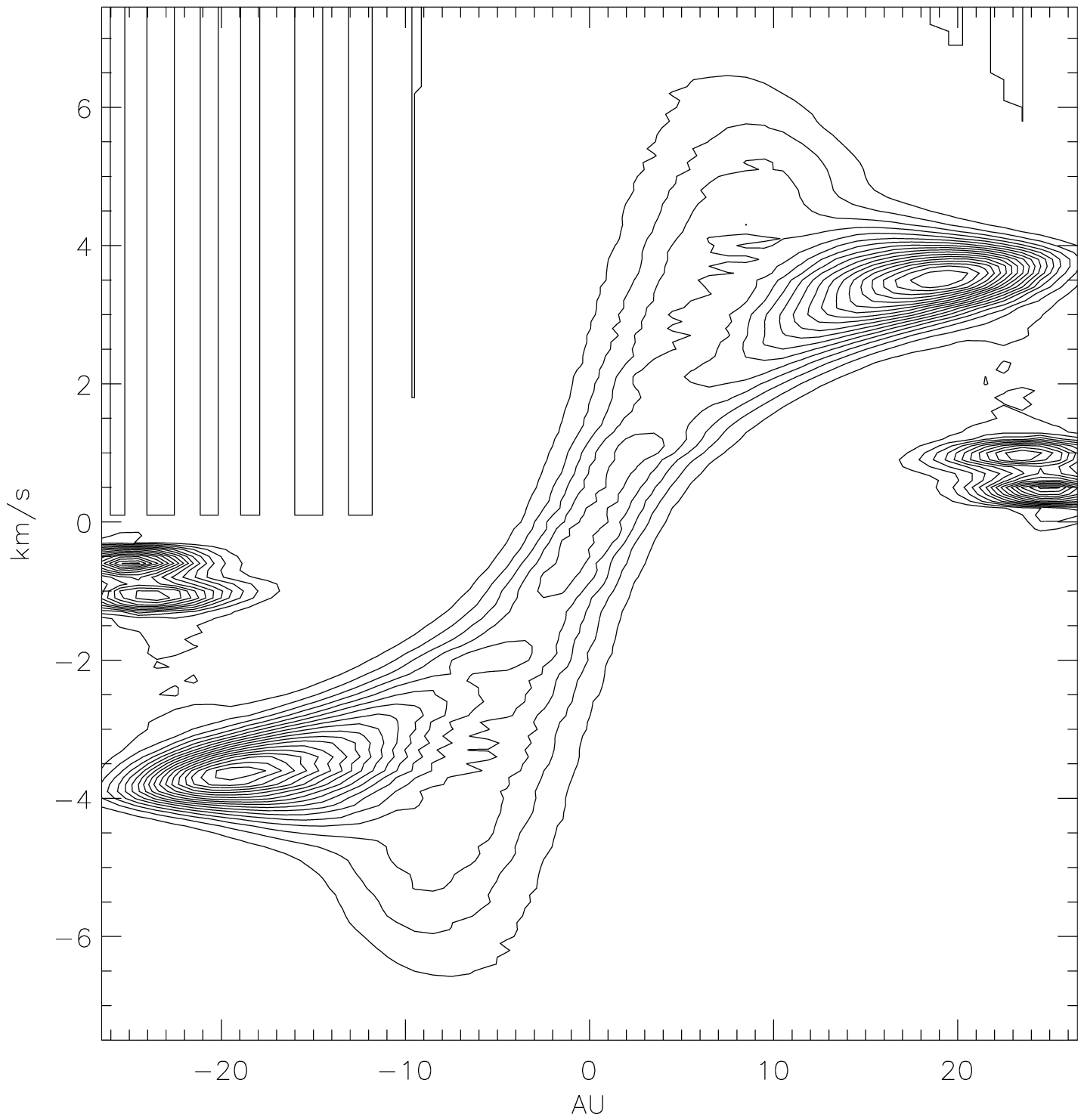}
\end{tabular}
\caption{Simulated \emph{ALMA} observation of a Jupiter mass planet in a \unit{10}{\AU} orbit.
The orbital plane is inclined at \unit{45}{\degree} to the line of sight, and the planet is located approximately at the 2 o'clock position.
The line of sight velocity is shown in the left panel, along with the beam track (which does not pass over the planet).
The circle placed on beam track indicates the \unit{5}{\AU} size of the \emph{ALMA} beam.
In the right panel, we show the position-velocity plot produced along the track}
\label{fig:JupiterALMAPlotsTrack1}
\end{figure}

\begin{figure}
\centering
\begin{tabular}{cc}
\includegraphics[width=4cm]{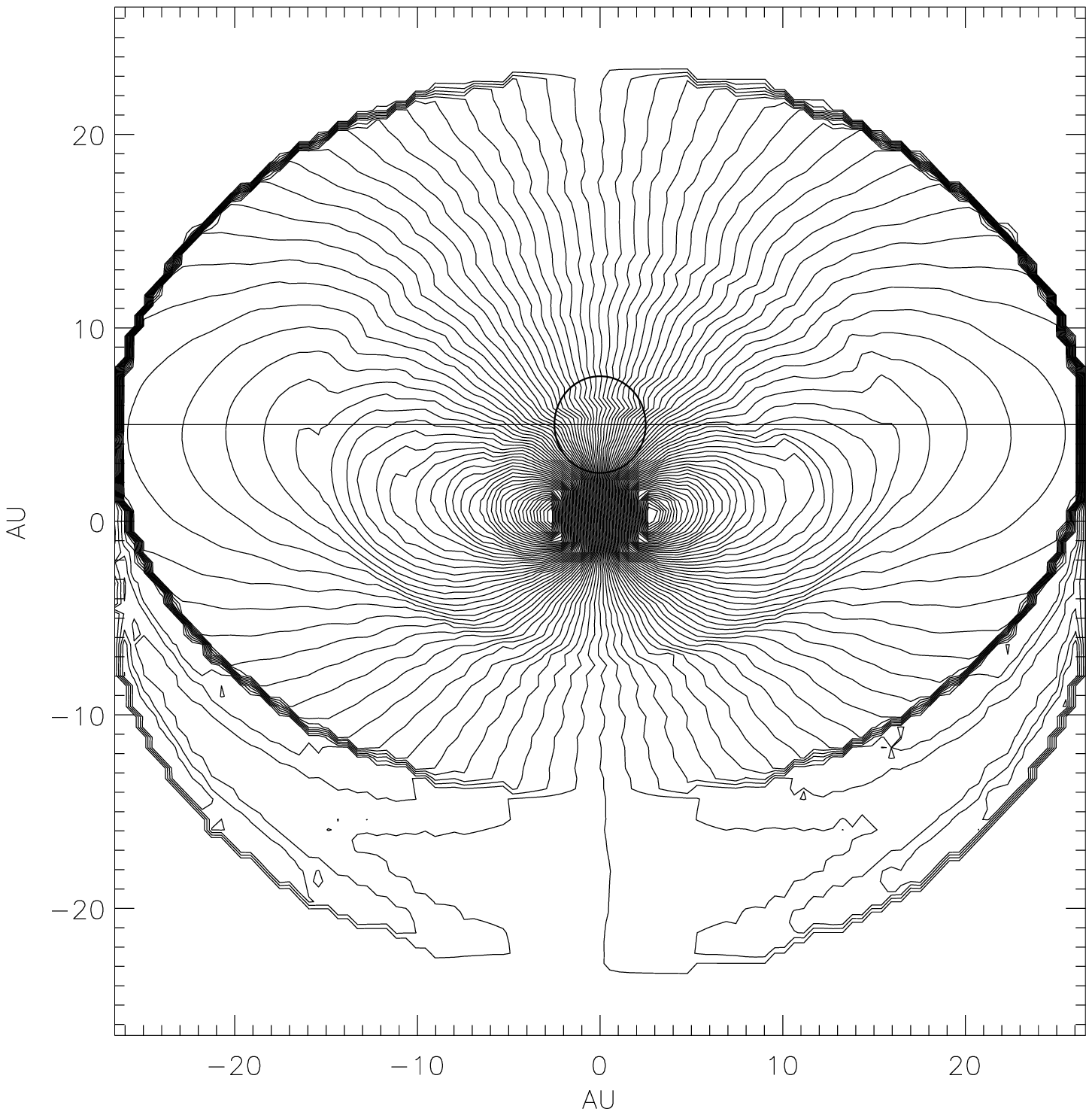} &
\includegraphics[width=4cm]{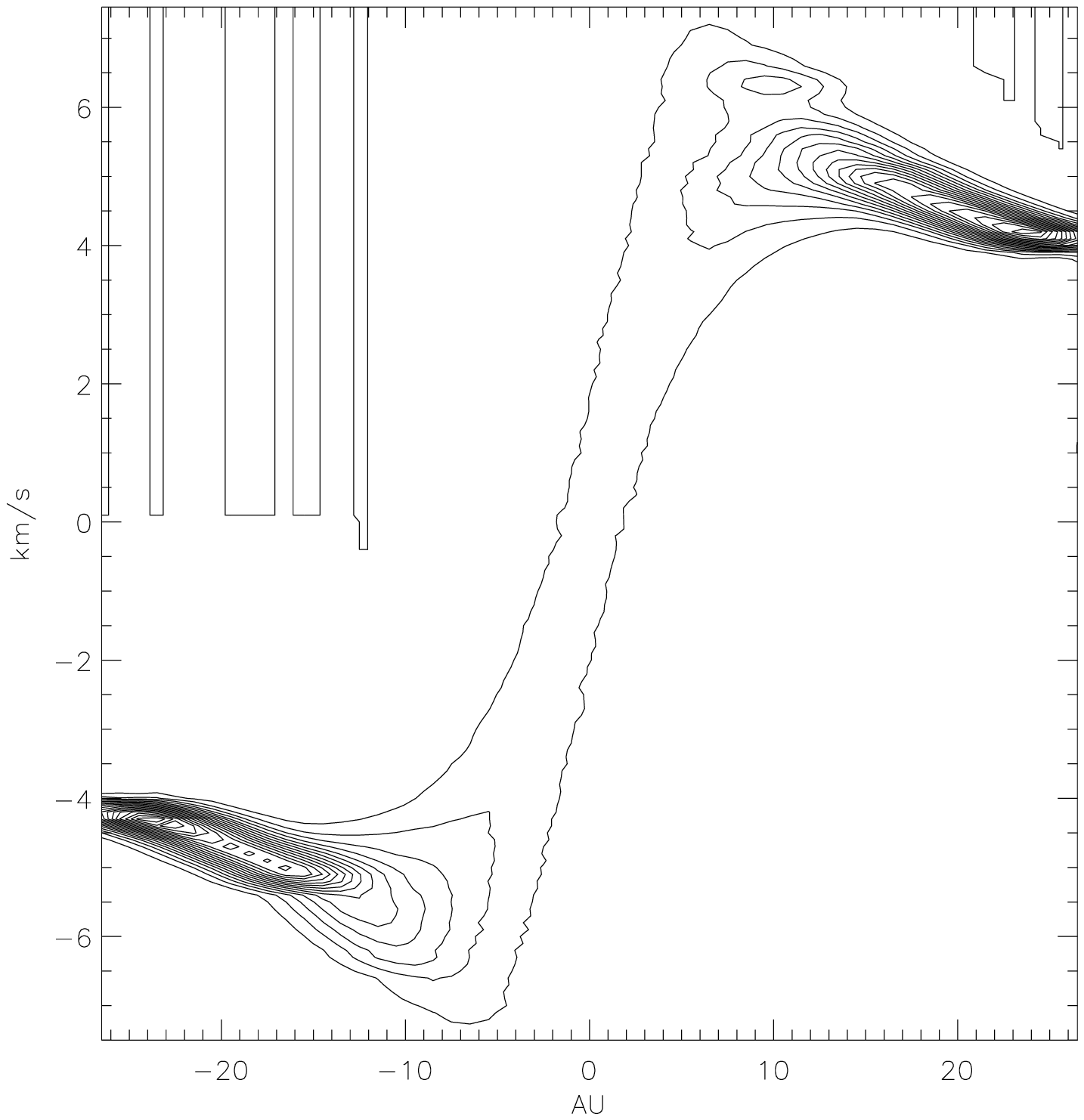}
\end{tabular}
\caption{Simulated \emph{ALMA} observation of a Jupiter mass planet in a \unit{10}{\AU} orbit.
The orbital plane is inclined at \unit{45}{\degree} to the line of sight, and the planet is located approximately at the 2 o'clock position.
The line of sight velocity is shown in the left panel, along with the beam track (which passes over the planet).
The circle placed on beam track indicates the \unit{5}{\AU} size of the \emph{ALMA} beam.
In the right panel, we show the position-velocity plot calculated for this beam track}
\label{fig:JupiterALMAPlotsTrack2}
\end{figure}

In Figures~\ref{fig:JupiterALMAPlotsTrack1} and~\ref{fig:JupiterALMAPlotsTrack2}, we show two simulated position-velocity plots, based on a Jupiter mass planet in the low viscosity disc.
The plots differ only in the beam track.
The disc was inclined at \unit{45}{\degree} to the line of sight, and we average the velocities over a \unit{5}{\AU} beam.
The planet was on the `far' side of the disc, in approximately the 2 o'clock position.
The beam track of Figure~\ref{fig:JupiterALMAPlotsTrack1} (the `near' side of the disc) does not pass over the planet, while that of Figure~\ref{fig:JupiterALMAPlotsTrack2} does.
Note that there is a clear difference in velocity structure between the `near' and `far' tracks, indicating that we can easily identify which side of the disc we are inspecting.

Of greater interest, is the fact that the planet is detectable in the right hand panel of Figure~\ref{fig:JupiterALMAPlotsTrack2}, as the extra contours around $x = \unit{10}{\AU}$ and $y = \unit{6}{\kilo\metre\usk\reciprocal\second}$.
This is slightly outside the planet's actual location (after projecting onto the sky), indicating that the detection is of the spiral arm raised by the planet.
Of course, real observations will be noisy, and the detection in Figure~\ref{fig:JupiterALMAPlotsTrack2} is only in one contour.
Furthermore, our approach of using the $\Sigma_{\text{los}}=1$ surface to construct our velocity field is crude.
Real observations will have to select particular molecular transitions, based on the temperature structure of the disc.
Optically thin lines will convolve velocities from many positions along the line of sight.

The vertical motions we have discovered might even make a planet detectable in a face-on disc, even if it is poorly resolved.
In this case, the vertical motions induced by the (unresolved) spiral arms might well be (mis-)interpreted as turbulent motions.
According to $\alpha$-disc theory, the turbulent velocity is $v_{\text{turb}} \approx \alpha^{1/2} c_s$.
If the disc were assumed to have a `normal' value of $\alpha \approx 10^{-3}$, then the sound speed in the disc would be over-estimated by a factor of $\mathcal{M}_z / \alpha^{1/2} \approx 10$, implying that the temperature would be over-estimated by a factor of roughly three.
At this point, we would only have a disc with an unreasonably large aspect ratio.
However, the temperature of the disc might be independently computable from observations of disc chemistry.
If these observations suggested a more reasonable temperature, then one might conclude that a planet was stirring up the disc, causing the mismatch between the `turbulent' temperature (in reality, the spiral arm stirring) and the chemical temperature.
Instruments such as the Submillimeter Array (SMA) are already making detailed observations of circumstellar discs \citep[e.g.][]{2004ApJ...616L..11Q,2006ApJ...636L.157Q}, and analysis of these systems may well find anomalies which can be explained by the presence of a planet.

Despite the caveats raised above, we have demonstrated the tantilising possibility of detecting a young planet embedded in is nascent disc.
Such a detection would greatly extend our understanding of the planet formation process.
Observational considerations have restricted all previous planet searches to old systems, where the disc has long since vanished.

%% file: conclude.tex
\section{Conclusions}
\label{sec:conclude}

In this paper, we have presented the results of \textsc{3d} numerical experiments of disc--planet interactions.
We have concentrated on the case of $1$ and \unit{2}{\Mjup} planets.
Our computations have shown strong vertical stirring of disc material by the planet.
This stirring has a number of potentially observable consequences.

The gap edge is likely to appear relatively thick for a number of reasons.
Direct illumination from the star will warm the gap edge, slightly increasing the scale height.
Density enhancements in the spiral arms might also be intepreted as a larger scale height in an unresolved image.
However, it is the coupling between the gas and dust which is most significant.
Vertical stirring by the planet's spiral arms will counteract dust sedimentation in the disc.
Combined, these effects will make the region close to the planet appear ``puffed up'' relative to the disc.
We argue that this strengthens the case for a planetary cause of the hole seen in CoKuTau/4, since the disc edge is not only sharp, but also quite thick.

We have also examined the possibility of detecting the velocity structure produced by the planet using the latest generation of telescope, such as \emph{ALMA}.
Rescaling the orbit of our planet to \unit{10}{\AU}, we believe that the presence of a planet is just detectable.
Even if the disc were unresolved, the effects of a planet might still be detectable.
A high `turbulent' temperature in a disc with a cold `chemistry' temperature would suggest the presence of a planet.

There are a number of avenues for further work, which we are actively pursuing.
Firstly, we are expanding our library of models, testing planets of different mass, in discs of varying viscosity and aspect ratio.
We have demonstrated that planet-induced velocity structures should be detectable, but we are not yet in a position to interpret real observations.
Compiling a library of models will provide us with the tools necessary to interpret future observations in detail.
Our simulated observations are somewhat crude, since they take no account of disc chemistry (which directly affects which molecules are available for observation; freeze-out of CO could be a problem).
We will refine our models in the future, computing the disc temperature and chemistry in more detail, and comparing these to the temperature profile assumed by \textsc{Flash}.

We have shown how planets embedded in circumstellar discs create significant features in the vertical structure of those discs.
These structures were not visible in earlier \textsc{2d} computations.
The vertical features created are likely to have detectable consequences, raising the exciting possibility of catching planets as they form.

%% file: changegamma.tex
\section{Effect of the Equation of State}
\label{sec:changegamma}

All the numerical experiments presented above used an adiabatic gas with $\gamma=1.1$ in the Riemann solver.
However, we imposed a temperature structure, and did not evolve the energy equation.
This is not entirely self-consistent.
To check the effect of this inconsistency on our results, we re-ran the low viscosity case for a Jupiter mass planet, with $\gamma=1.01$.

\begin{figure}
\centering
\subfigure{\includegraphics[scale=0.6]{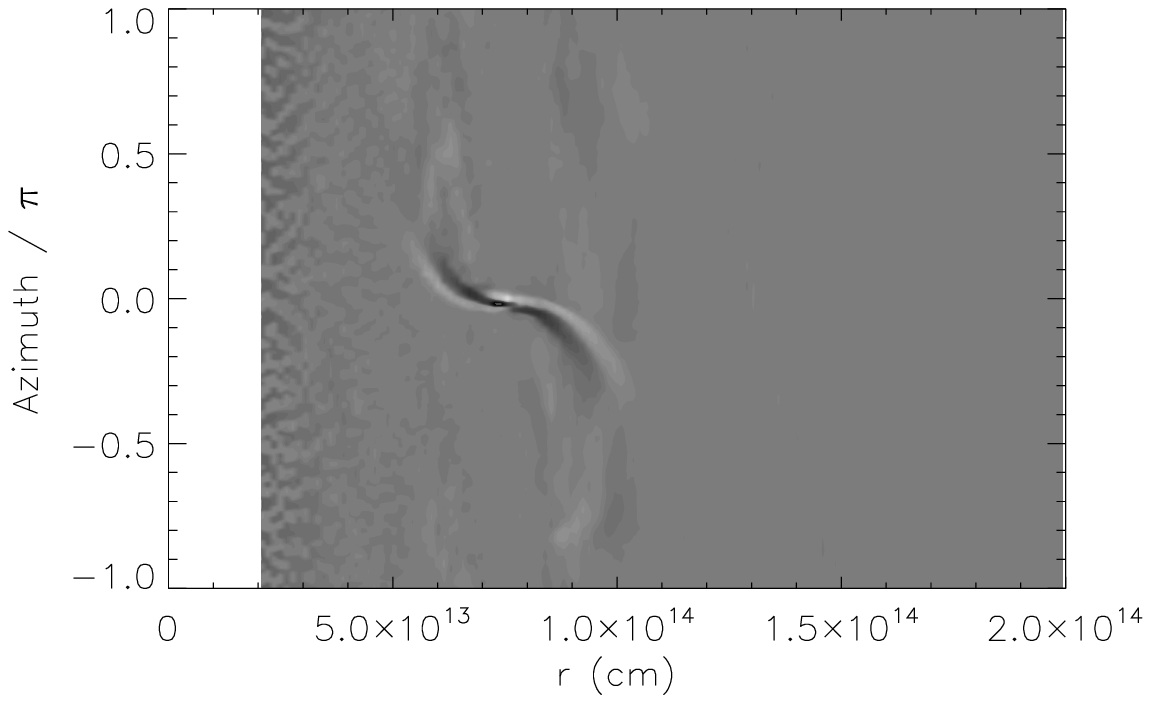}}
\subfigure{\includegraphics[scale=0.6]{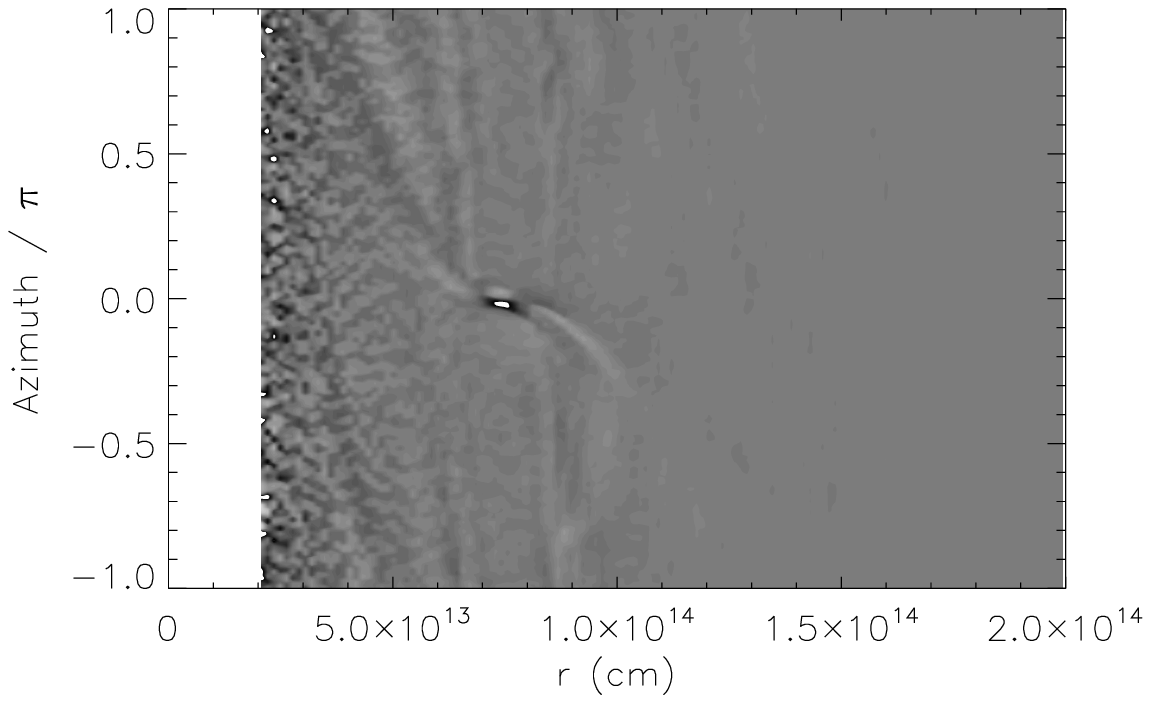}}
\includegraphics[scale=0.6]{MachBar}
\caption{The effect of the choice of $\gamma$ on the Mach number of the vertical velocity on planes with $z=\rHill{}/4$ (top) and $\rHill{}$ (bottom) for a $q=10^{-3}$ planet in the low viscosity disc. The scale ranges from $\mathcal{M}_{z}=-1$ (black) to 1 (white). The numerical experiment producing this Figure was identical to that of the middle row of Figure~\ref{fig:MachZSliceZJupiter}, except this Figure uses $\gamma=1.01$ in the Riemann solver}
\label{fig:ChangeGammaMachZSliceZLowVisc}
\end{figure}

In Figure~\ref{fig:ChangeGammaMachZSliceZLowVisc}, we show the effect of this change on the value of \Machz{} on slices with $z=\rHill{}/4$ (top) and \rHill{} (bottom).
Apart from the value of $\gamma$, the numerical experiment used to produce this Figure was identical to that of the middle row of Figure~\ref{fig:MachZSliceZJupiter}.
We see that similar behaviour occurs in the disc with a $\gamma=1.01$ gas.

Ideally, one should solve the energy equation in the disc \citep[e.g.][]{2003ApJ...599..548D}, perhaps even considering radiative transfer.
However, this test reassures us that the vertical motions induced by the planet are real, and not merely artifacts of the numerical method.

%% file: viscouscomponents.tex
\section{Adding Components of the Viscous Stress Tensor}
\label{sec:viscouscomponents}

As noted in Section~\ref{sec:numerics}, the runs presented above only used the \mbox{$r$--$\phi$} component of the viscous stress tensor.
We made this decision because the true nature of viscosity in accretion discs is not known, and we simply wished to include a term which would cause angular momentum transport.
However, our code implements all components of the viscous stress tensor, and in this section, we present a run where all components were included.

\begin{figure}
\centering
\subfigure{\includegraphics[scale=0.6]{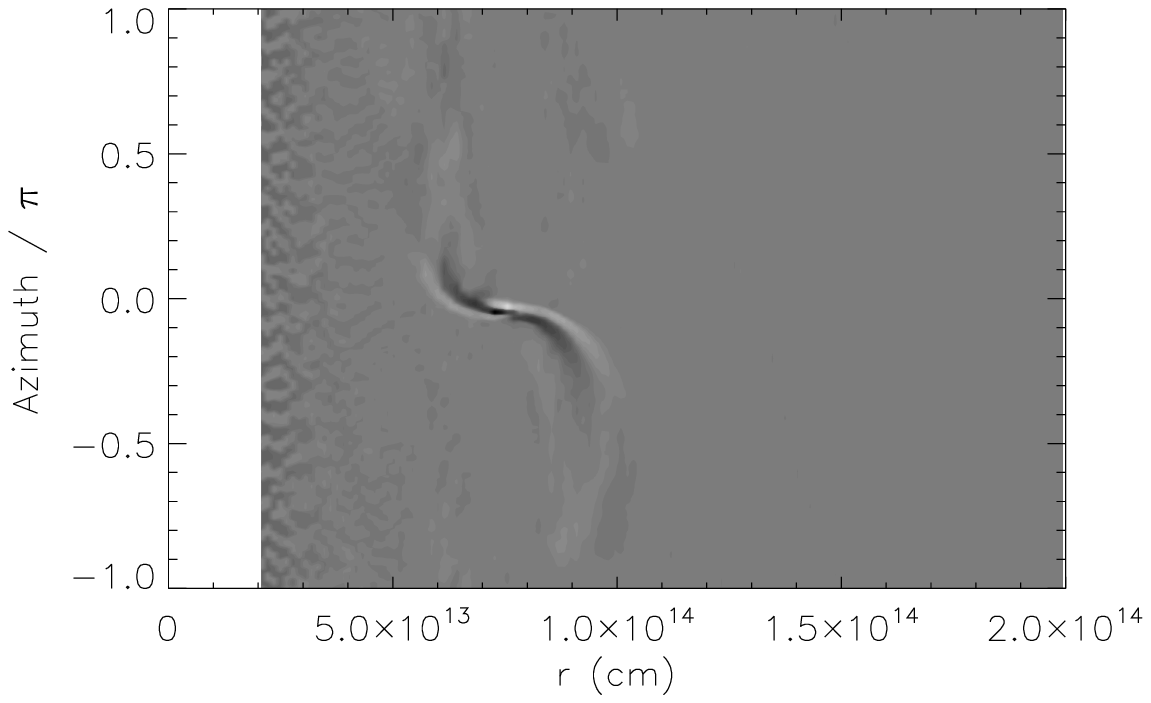}}
\subfigure{\includegraphics[scale=0.6]{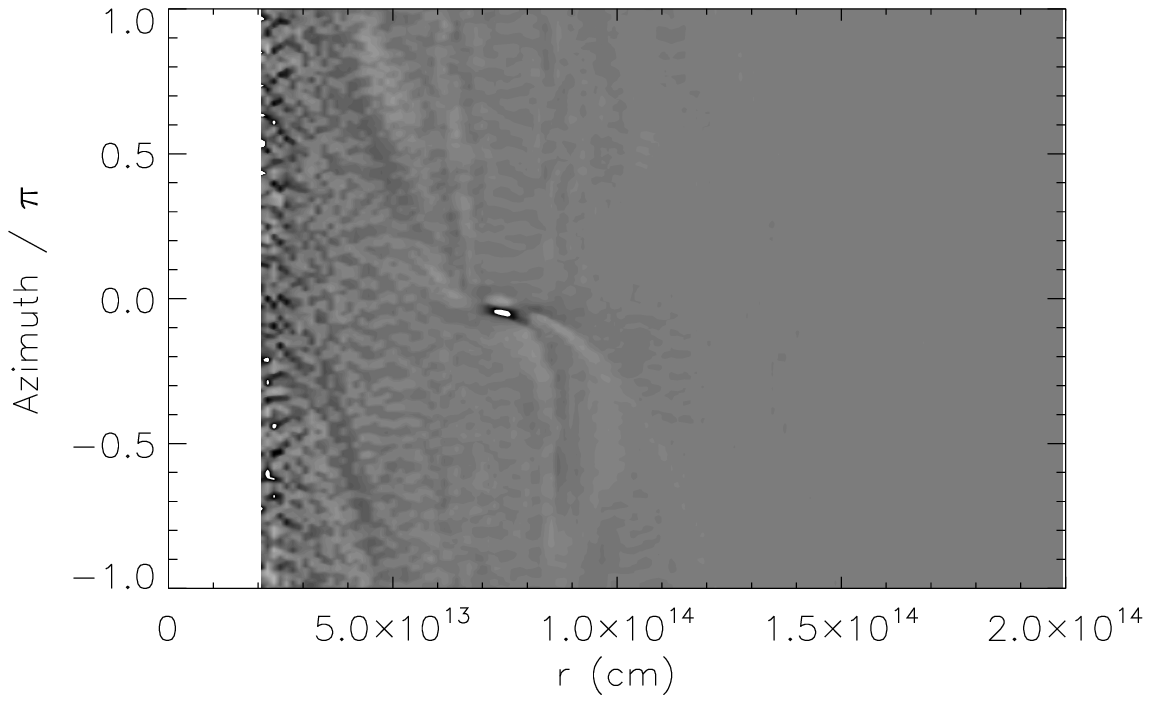}}
\includegraphics[scale=0.6]{MachBar}
\caption{The effect of including all components of the viscous stress tensor on the Mach number of the vertical velocity on planes with $z=\rHill{}/4$ (top) and $\rHill{}$ (bottom) for a $q=10^{-3}$ planet in the low viscosity disc. The scale ranges from $\mathcal{M}_{z}=-1$ (black) to 1 (white). The numerical experiment producing this Figure was identical to that of the middle row of Figure~\ref{fig:MachZSliceZJupiter}, except this Figure for the inclusion of the extra viscous terms}
\label{fig:ViscousComponentsMachZSliceZLowVisc}
\end{figure}

In Figure~\ref{fig:ViscousComponentsMachZSliceZLowVisc}, we show the effect of including all the components of the viscous stress tensor on the value of \Machz{} on slices with $z=\rHill{}/4$ (top) and \rHill{} (bottom).
Apart from the inclusion of these terms, the numerical experiment used to produce this Figure was identical to that of the middle row of Figure~\ref{fig:MachZSliceZJupiter}.
We can see that the inclusion of the extra viscous terms has not suppressed the strong vertical motions we saw previously.
As with the case of increased viscosity, this is not surprising: the viscosity affects the structure of the disc over many orbits, whereas the planet stirs the disc on its orbital timescale.
Viscosity can act to damp some of the fine structure, but it cannot prevent the planet exciting strong vertical motions within the disc.

%% file: resolution.tex
\section{Effect of Resolution}
\label{sec:resolution}

\begin{figure}
\centering
\subfigure{\includegraphics[scale=0.6]{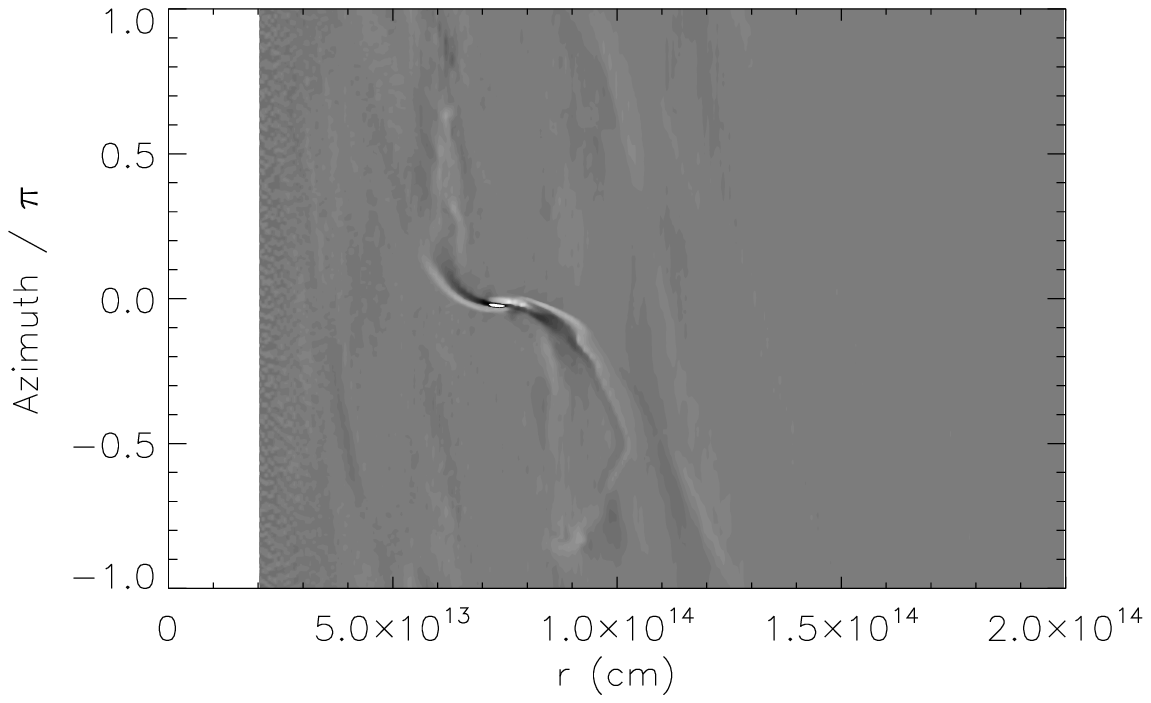}}
\subfigure{\includegraphics[scale=0.6]{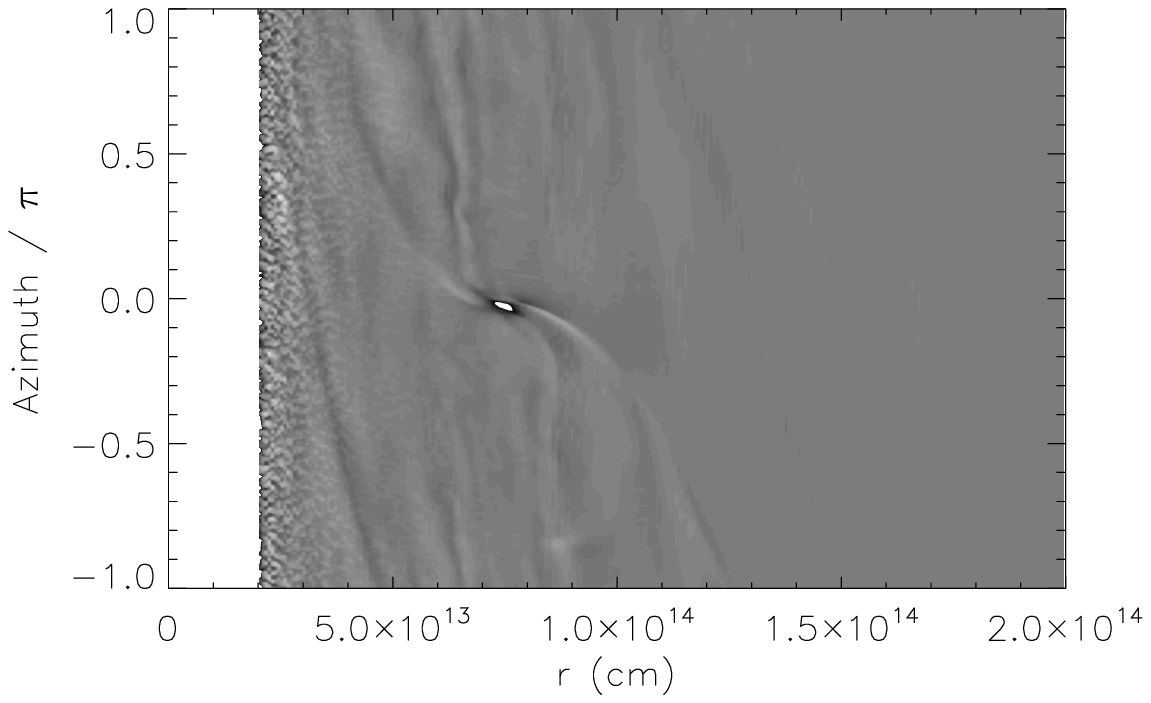}}
\includegraphics[scale=0.6]{MachBar}
\caption{The effect of doubling the grid resolution on the Mach number of the vertical velocity on planes with $z=\rHill{}/4$ (top) and $\rHill{}$ (bottom) for a $q=10^{-3}$ planet in the low viscosity disc. The scale ranges from $\mathcal{M}_{z}=-1$ (black) to 1 (white). This is a continuation of the numerical experiment which produced the middle row of Figure~\ref{fig:MachZSliceZJupiter}, with the resolution doubled}
\label{fig:DoubleResolutionMachZSliceZLowVisc}
\end{figure}

As noted in Section~\ref{sec:results}, our resolution (particularly the vertical resolution) in the innermost portions of the disc is not especially good.
This is a result of necessity: global \textsc{3d} numerical experiments are extremely computationally expensive.
However, it is important to verify that numerical results are not purely artifacts of resolution.
Accordingly, we instructed \textsc{Flash} to continue the run with Jupiter in a low viscosity disc, but with the resolution doubled.
Figure~\ref{fig:DoubleResolutionMachZSliceZLowVisc} shows the result of this continuation.

We can see that the structure in \Machz{} is retained.
However, the noise in the inner portions of the disc has been substantially suppressed (albeit not eradicated), especially in the $z=\rHill{}$ plane.
The peaks in \Machz{} appear to be slightly smaller in Figure~\ref{fig:DoubleResolutionMachZSliceZLowVisc}, indicating that our runs at standard resolution might be slightly under-resolving the vertical stirring.
This said, the structures seen are very similar.